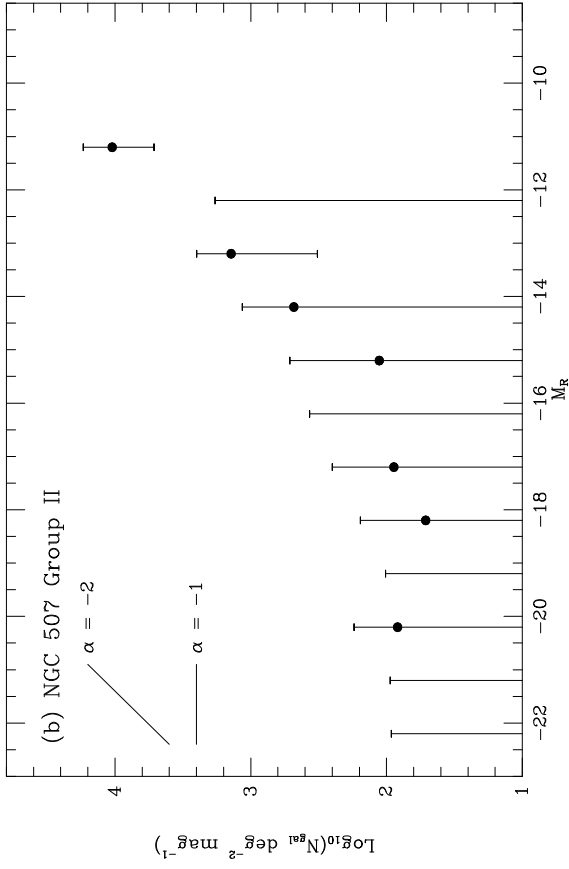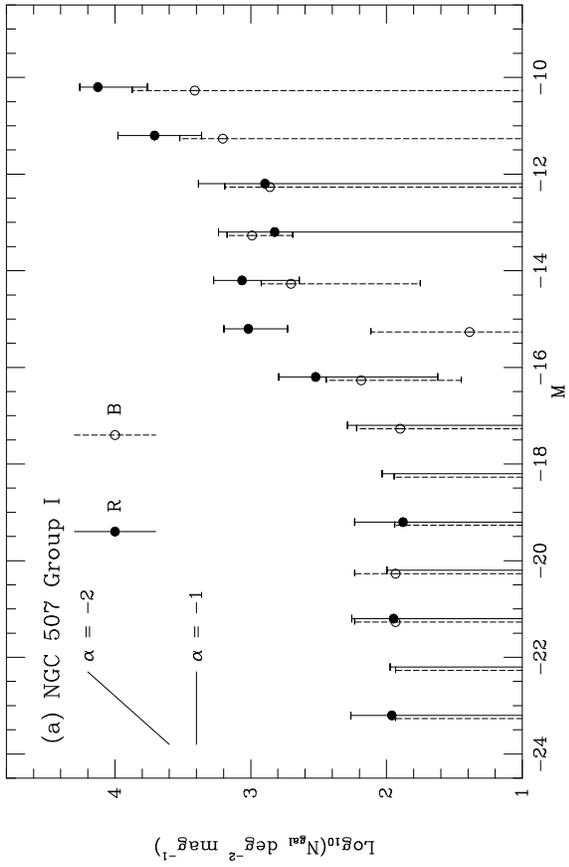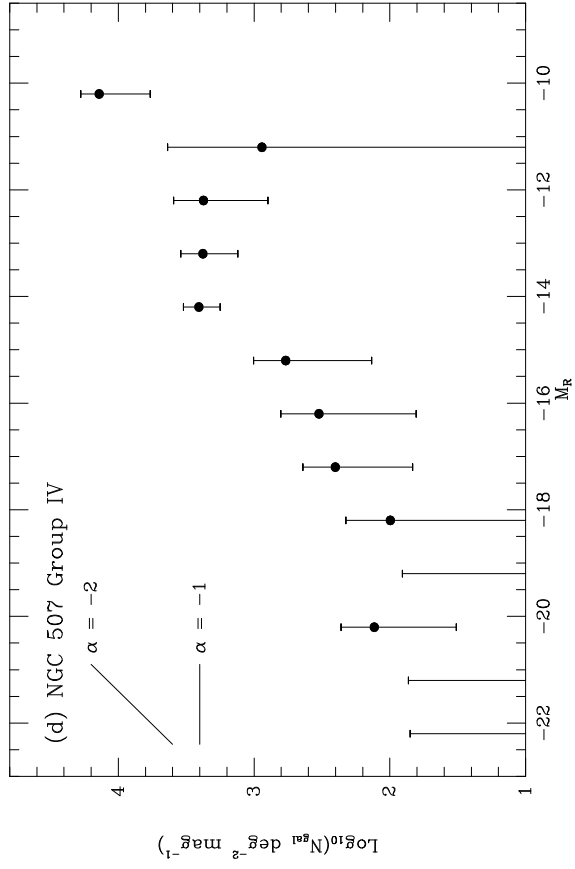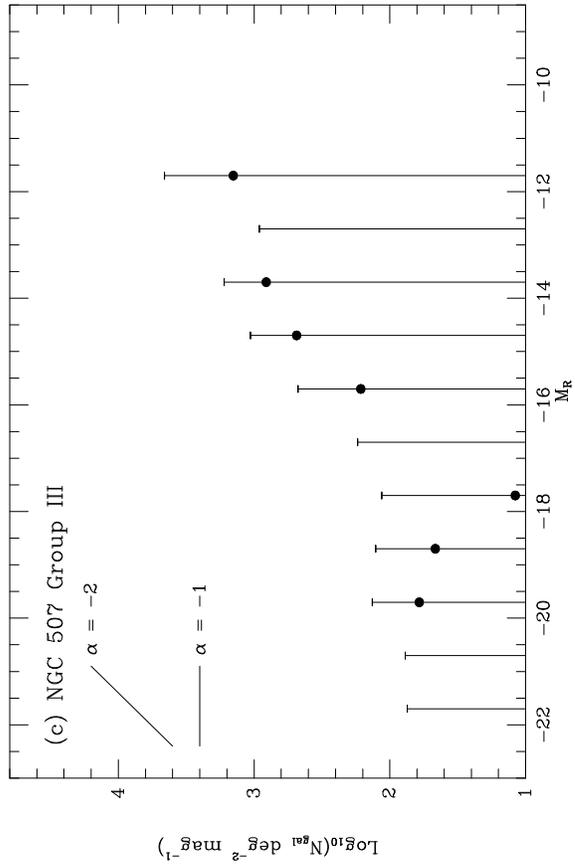

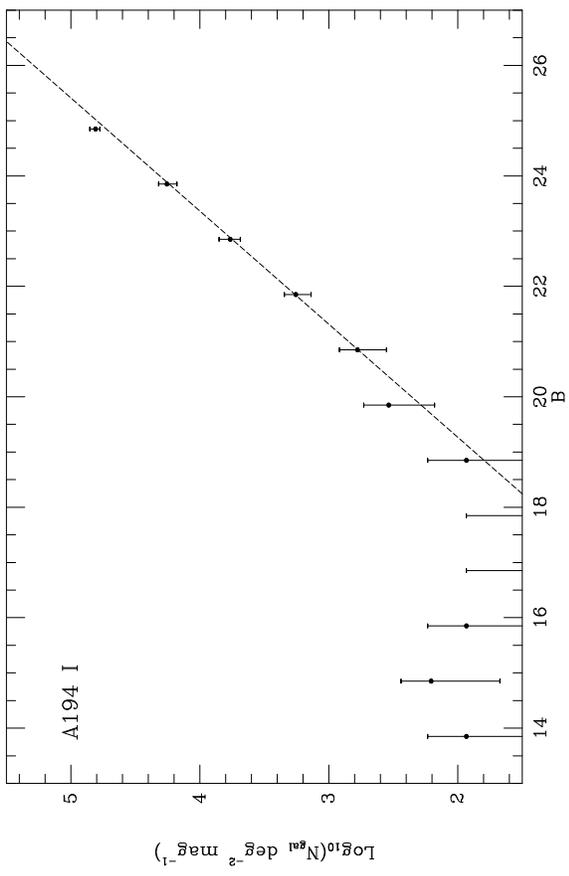
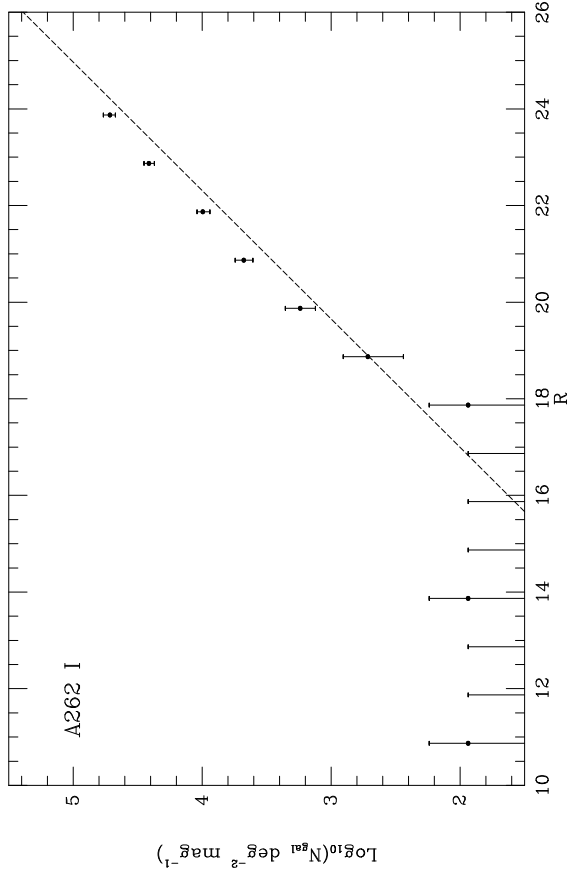
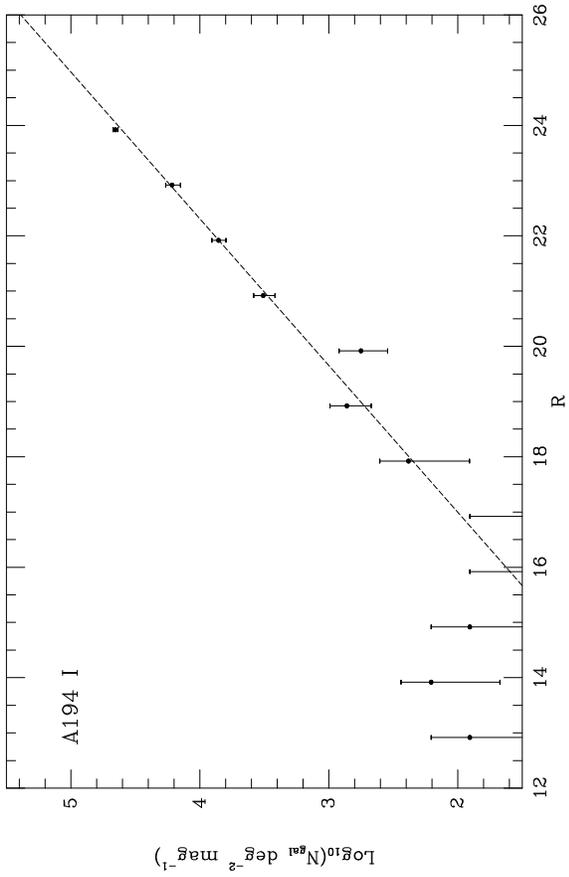
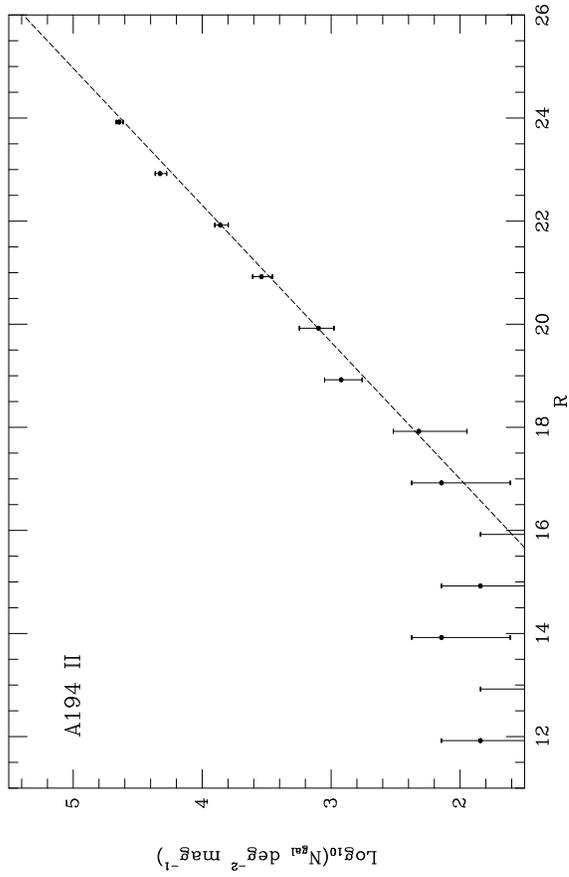

Table 1

Properties of the Sample Clusters

| Cluster | Richness (1) | $z$ | $\sigma$ km s$^{-1}$ | $L_x$ erg s$^{-1}$ | $A_B$ (8) mag | $\Sigma_c$ g cm$^{-2}$ |
|---|---|---|---|---|---|---|
| A194 | 0 | 0.018 | 440 (2) | $0.1 \times 10^{43}$ (5)* | 0.15 | 4.9 |
| A262 | 0 | 0.016 | 506 (3) | $1.4 \times 10^{43}$ (5)* | 0.24 | 5.5 |
| Pegasus | < 0 | 0.014 | 639 (4) | $< 0.3 \times 10^{43}$ (6)* | 0.16 | 6.3 |
| NGC 507 Group | < 0 | 0.016 | 511 (3) | $0.4 \times 10^{43}$ (7)† | 0.17 | 5.5 |

* $0.5 - 3.0$ keV Luminosity

† $0.2 - 4.0$ keV Luminosity

**References:** (1) Abell 1958; (2) Chapman et al. 1988; (3) Sakai et al. 1994; (4) Bothun et al. 1985; (5) Jones & Forman 1984; (6) Giovanelli & Haynes 1985; (7) Kim & Fabbiano 1995; (8) Burstein & Heiles 1982.

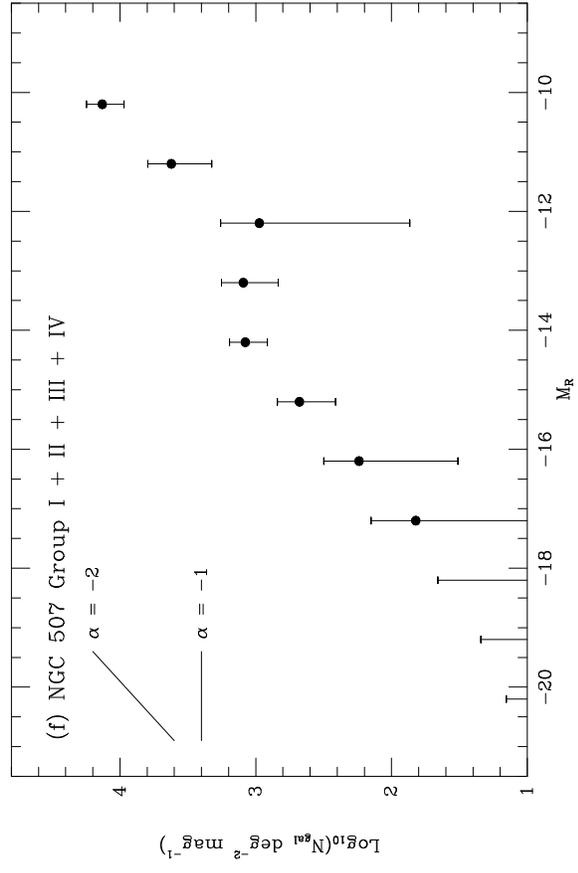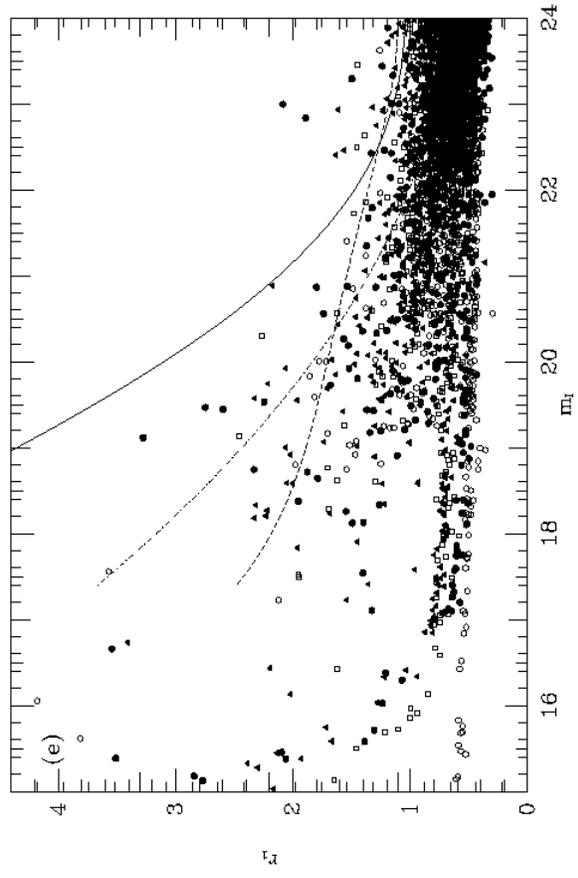

a)

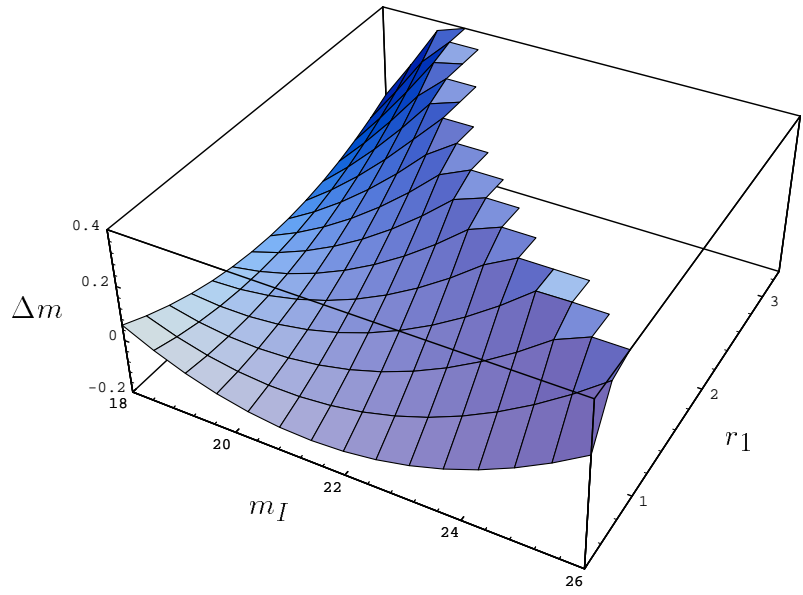

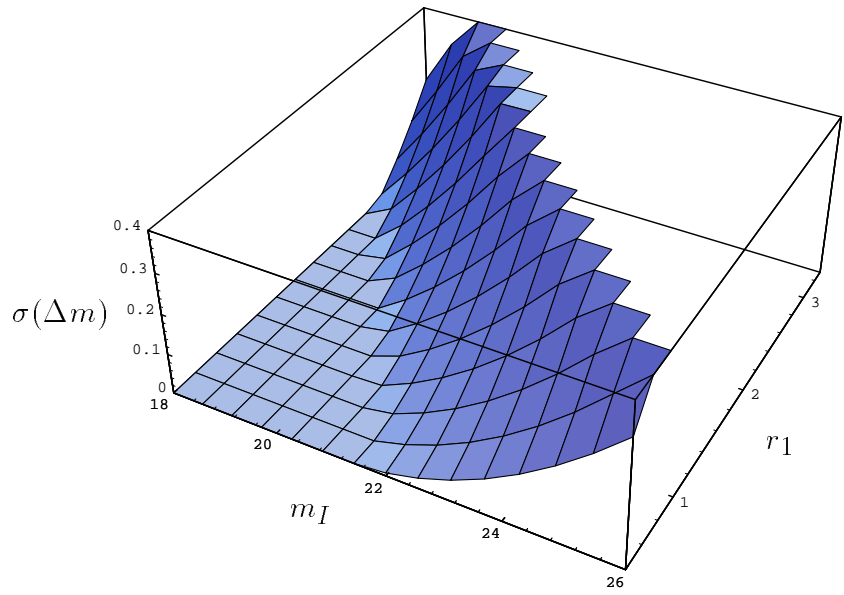

b)

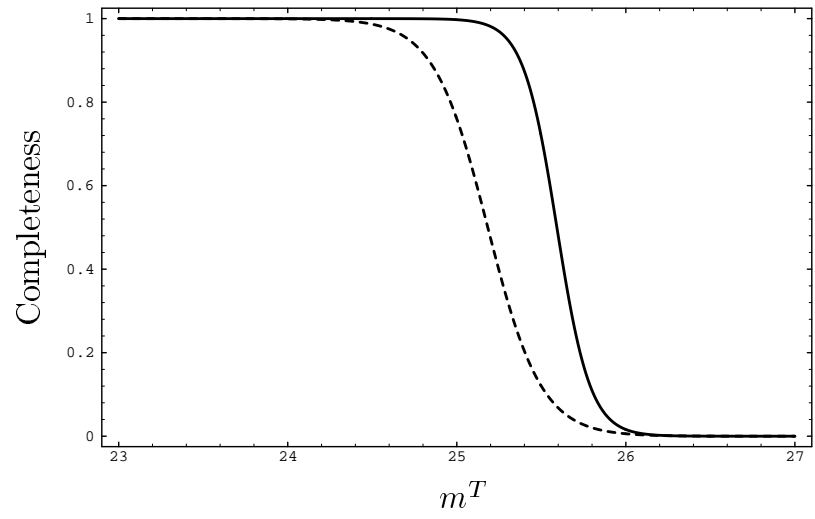

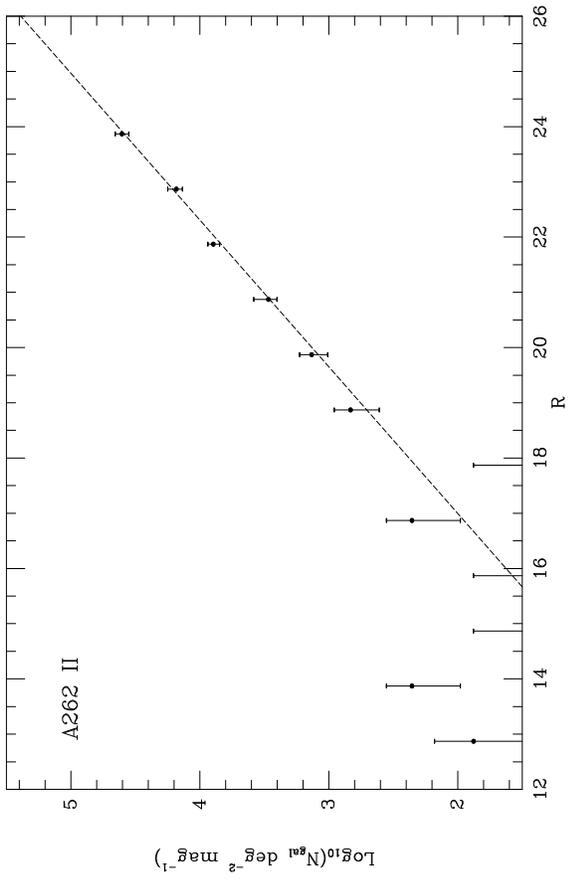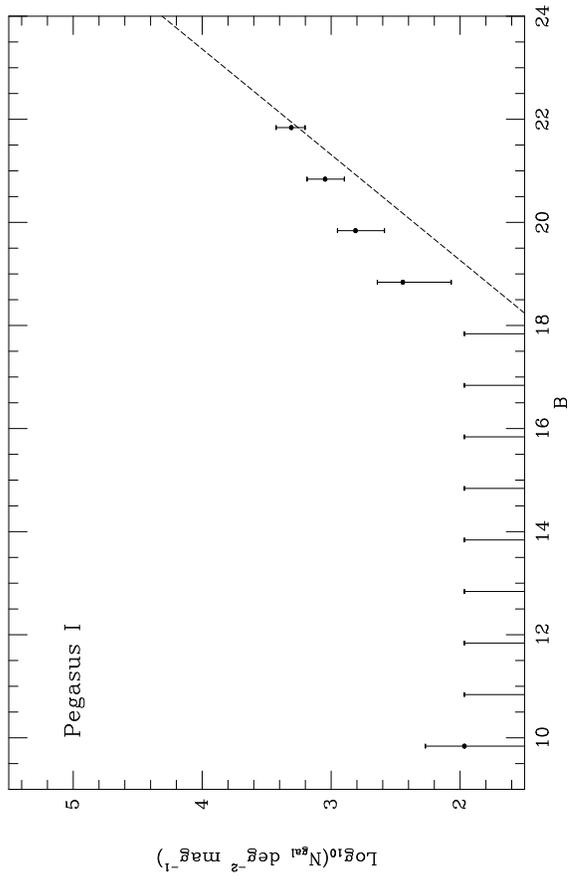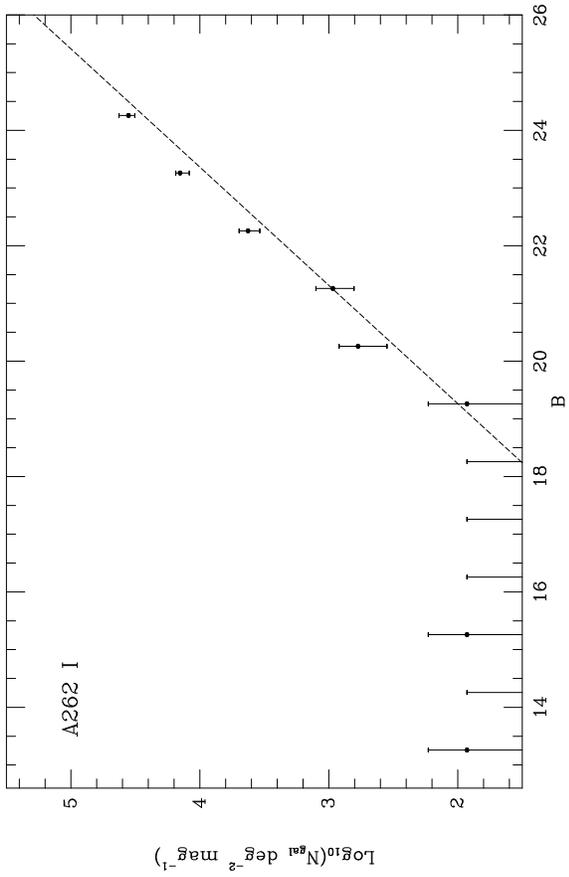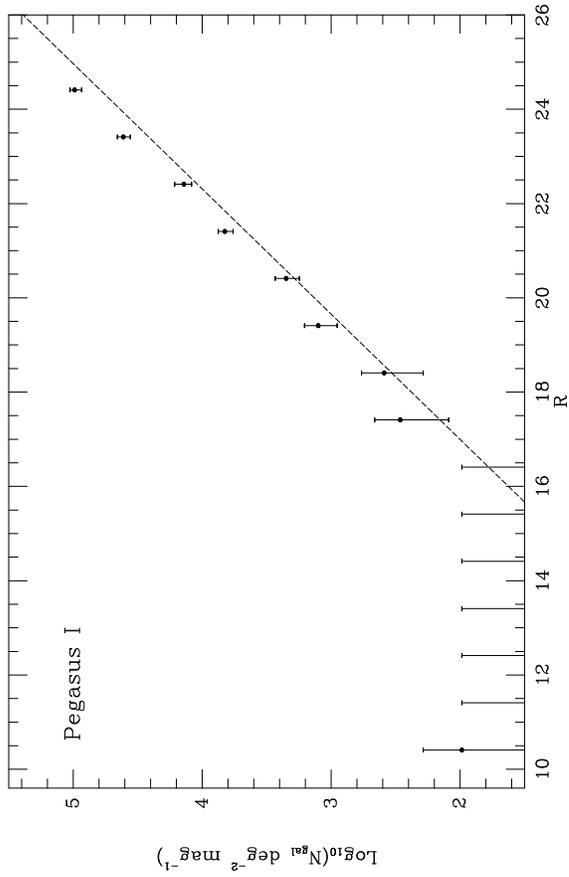

## Table 2
## Observing Log

| Field | Color | Run* | Size (sq. arcmin) | $\alpha$ (1950) | $\delta$ (1950) | Exposure min | $<X>$ mag | FWHM arcsec |
|---|---|---|---|---|---|---|---|---|
| A194 I | R | 3 | 46.5 | $1^h23.0^m$ | $-1°46.0'$ | 60 | 0.00 | 0.79 |
|  | B | 5 | 45.4 |  |  | 140 | 0.37 | 0.88 |
| A194 II | R | 3 | 56.0 | $1^h22.0^m$ | $-1°39.2'$ | 60 | 0.00 | 0.95 |
| A262 I | R | 3 | 46.1 | $1^h50.0^m$ | $35°55.0'$ | 80 | 0.39 | 0.85 |
|  | B | 4 | 46.1 |  |  | 80 | 0.00 | 1.05 |
| A262 II | R | 3 | 54.1 | $1^h49.4^m$ | $35°55.0'$ | 60 | 0.00 | 0.92 |
| Pegasus I | R | 3 | 43.7 | $23^h18.0^m$ | $7°55.1'$ | 60 | 0.10 | 0.83 |
|  | B | 4,5 | 43.7 |  |  | 100 | 0.28 | 0.92 |
| Pegasus II | R | 3 | 54.5 | $23^h18.0^m$ | $8°02.0'$ | 60 | 0.10 | 1.10 |
| N507 Group I | R | 3 | 44.0 | $1^h21.0^m$ | $33°00.0'$ | 60 | 0.12 | 0.81 |
|  | B | 5 | 44.0 |  |  | 160 | 0.74 | 0.89 |
| N507 Group II | R | 3 | 54.5 | $1^h20.4^m$ | $33°00.0'$ | 60 | 0.55 | 1.01 |
| N507 Group III | R | 5 | 54.5 | $1^h21.0^m$ | $33°07.0'$ | 60 | 1.00 | 0.76 |
| N507 Group IV | R | 3 | 45.7 | $1^h21.0^m$ | $33°07.0'$ | 60 | 0.03 | 0.99 |
| Background 1 | R | 1,2 | 42.7 | $10^h16.1^m$ | $51°07.4'$ | 90 | 0.00 | 0.89 |
|  | B | 1,2 | 42.7 |  |  | 55 | 0.00 | 0.89 |
| Background 2 | R | 3 | 55.7 | $17^h00.0^m$ | $30°00.0'$ | 50 | 0.00 | 0.82 |
| Background 3 | R | 6 | 89.7 | $10^h14.2^m$ | $37°06.8'$ | 60 | 0.00 | 1.47 |
|  | B | 6 | 89.7 |  |  | 60 | 0.00 | 1.42 |

***Observing runs: 1.** March 7−9 1994, **2.** April 8−10 1994, **3.** September 4−5 1994, **4.** November 3−5 1994, **5.** November 23−25 1994 (all UH 2.2 m with Tektronix 2048 × 2048 CCD at f/10); **6.** January 29−31 1995 (with MOS in direct imaging mode on the CFHT at f/8).

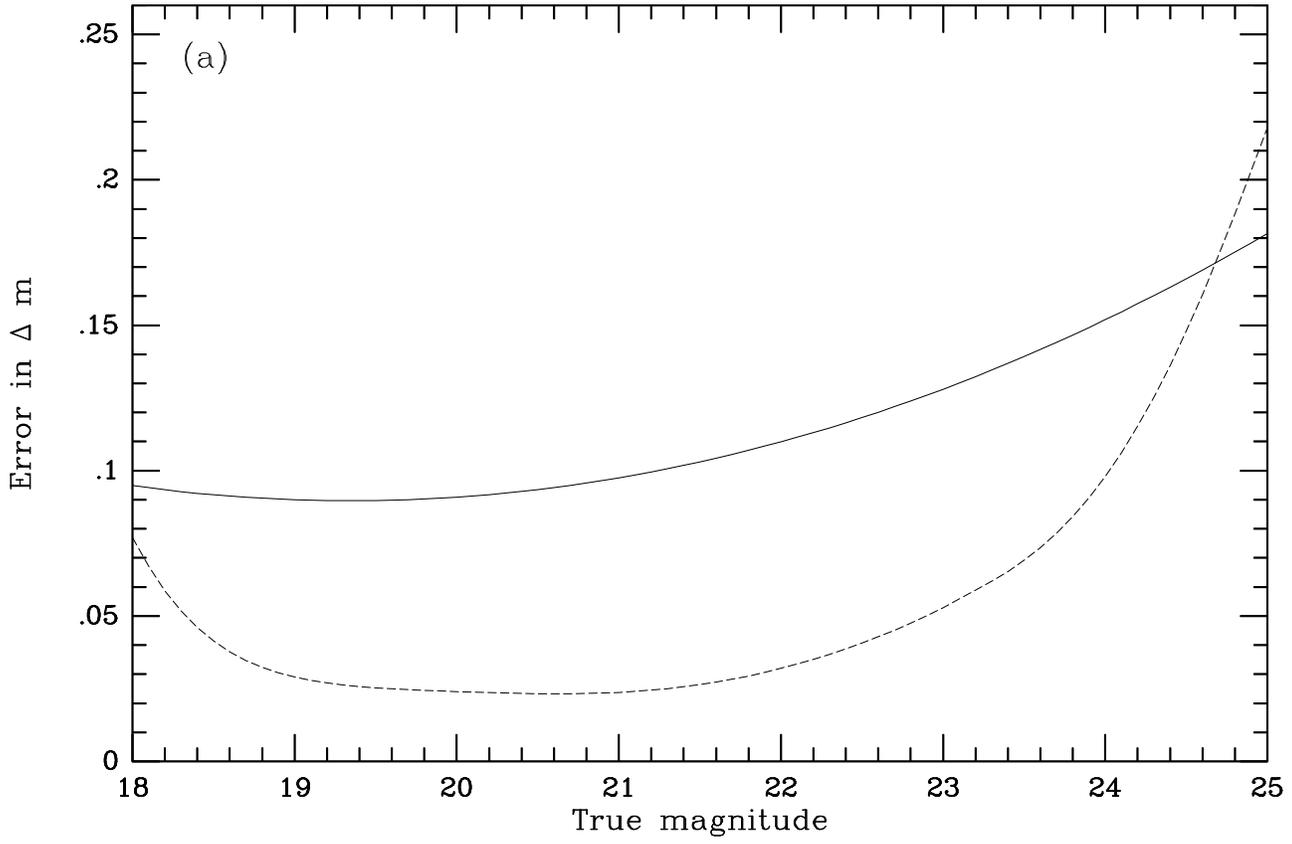

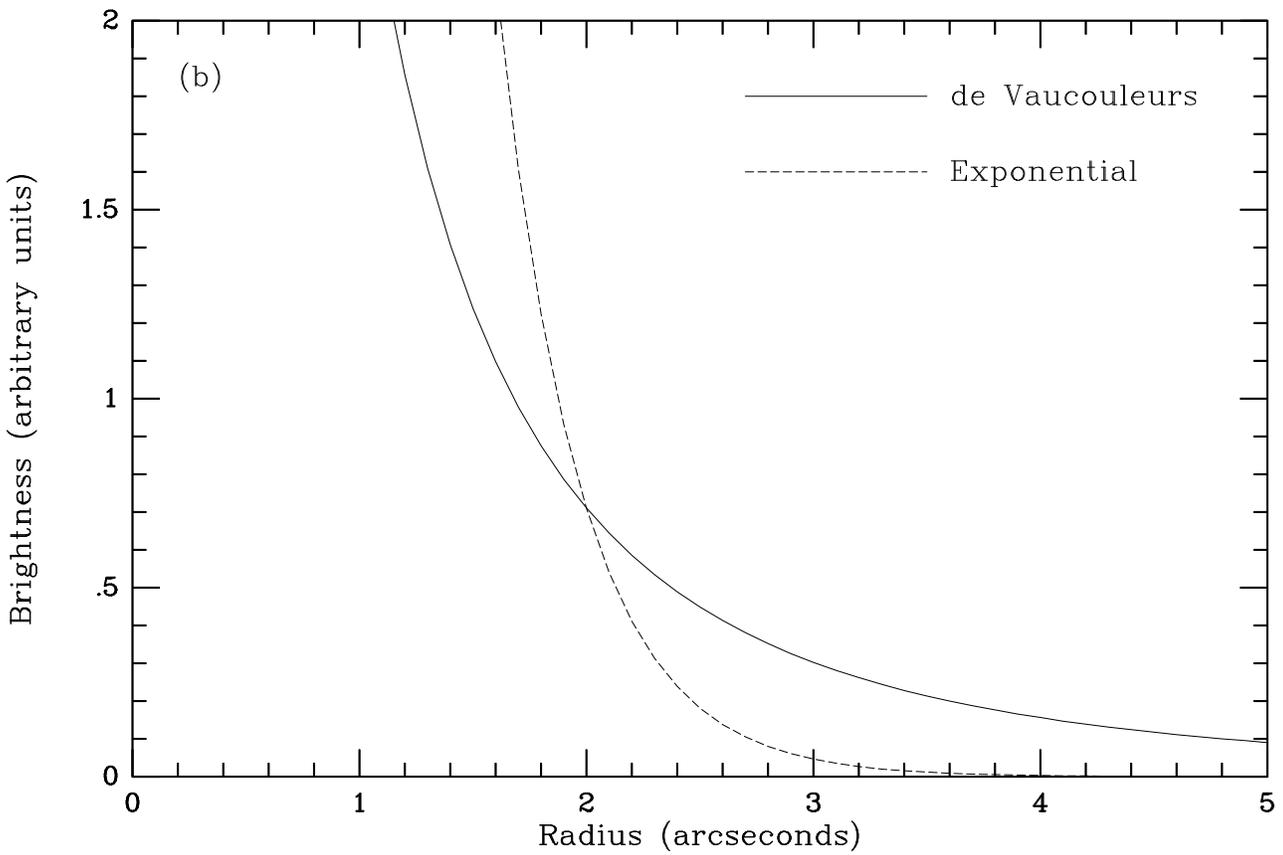

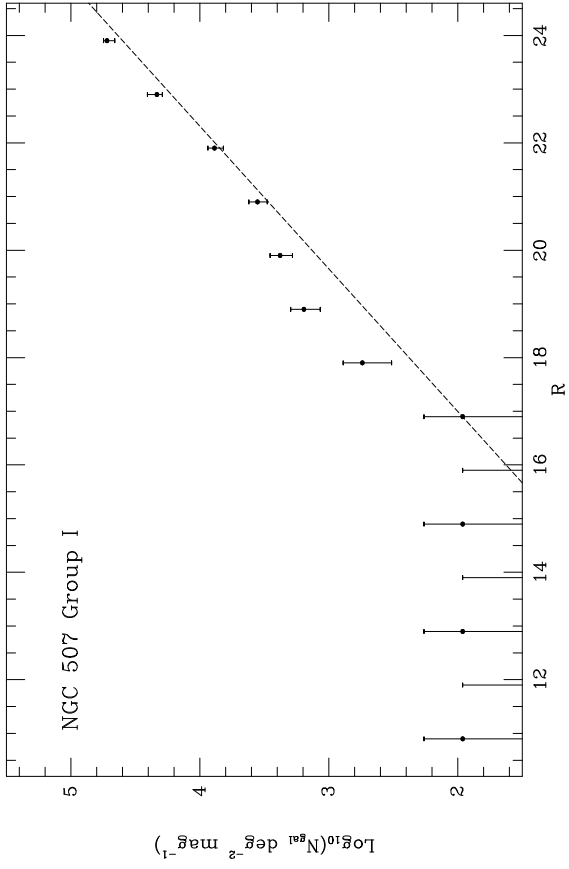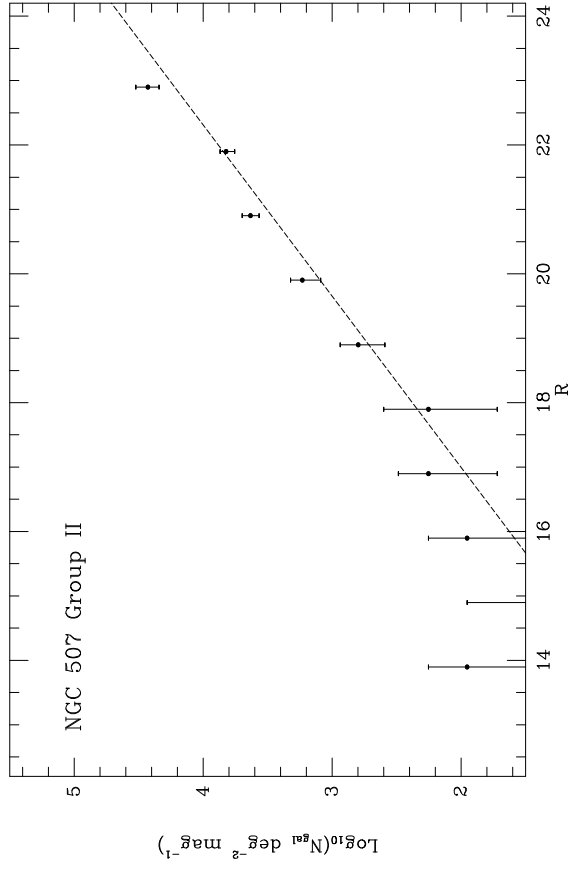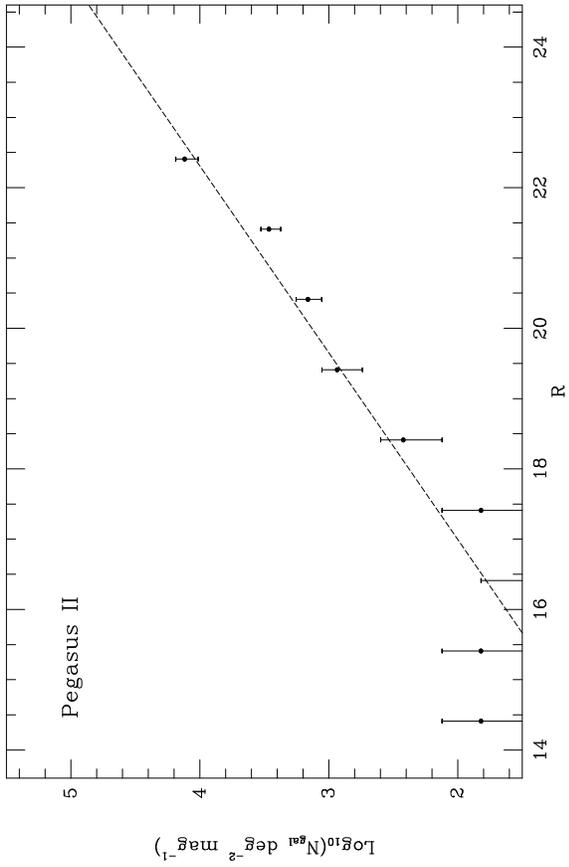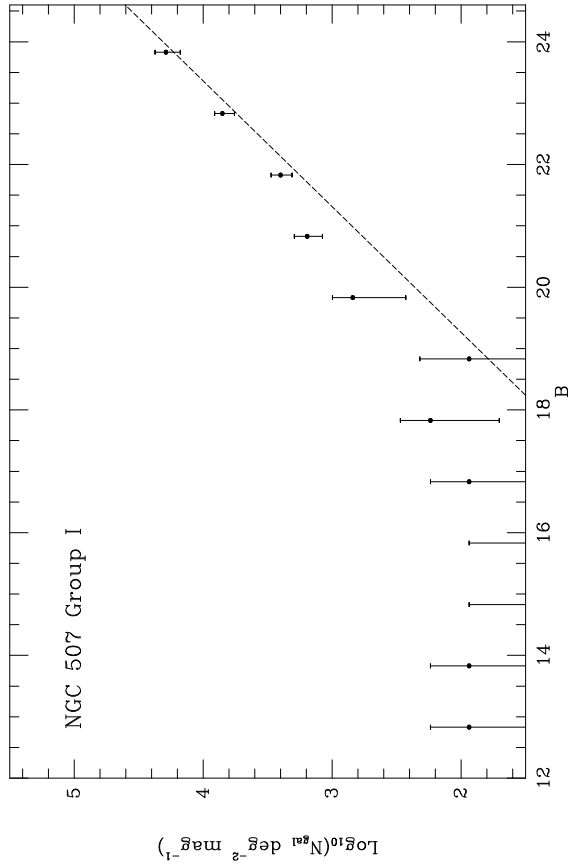

Table 3

Abell 194 Luminosity Function

| | $-15.88 < M_R < -9.88$ |
|---|---|
| $\eta = 1$ | $\alpha = -2.34^{+0.75}_{-\infty}$ |
| $\eta = 2$ | $\alpha = -2.25$ [$\alpha$ unconstrained at the $1\sigma$ confidence level] |
| $\eta = 3$ | $\alpha = -2.23$ [$\alpha$ unconstrained at the $1\sigma$ confidence level] |

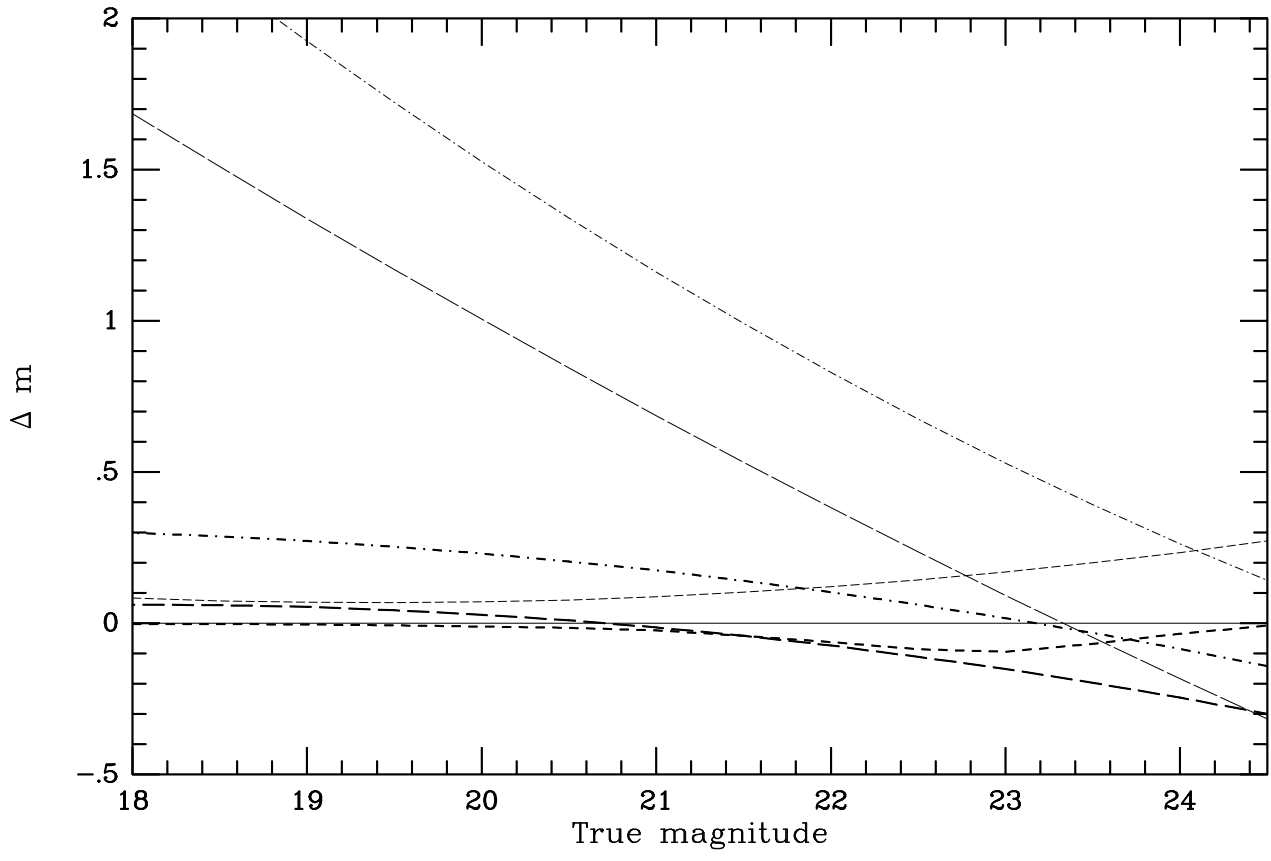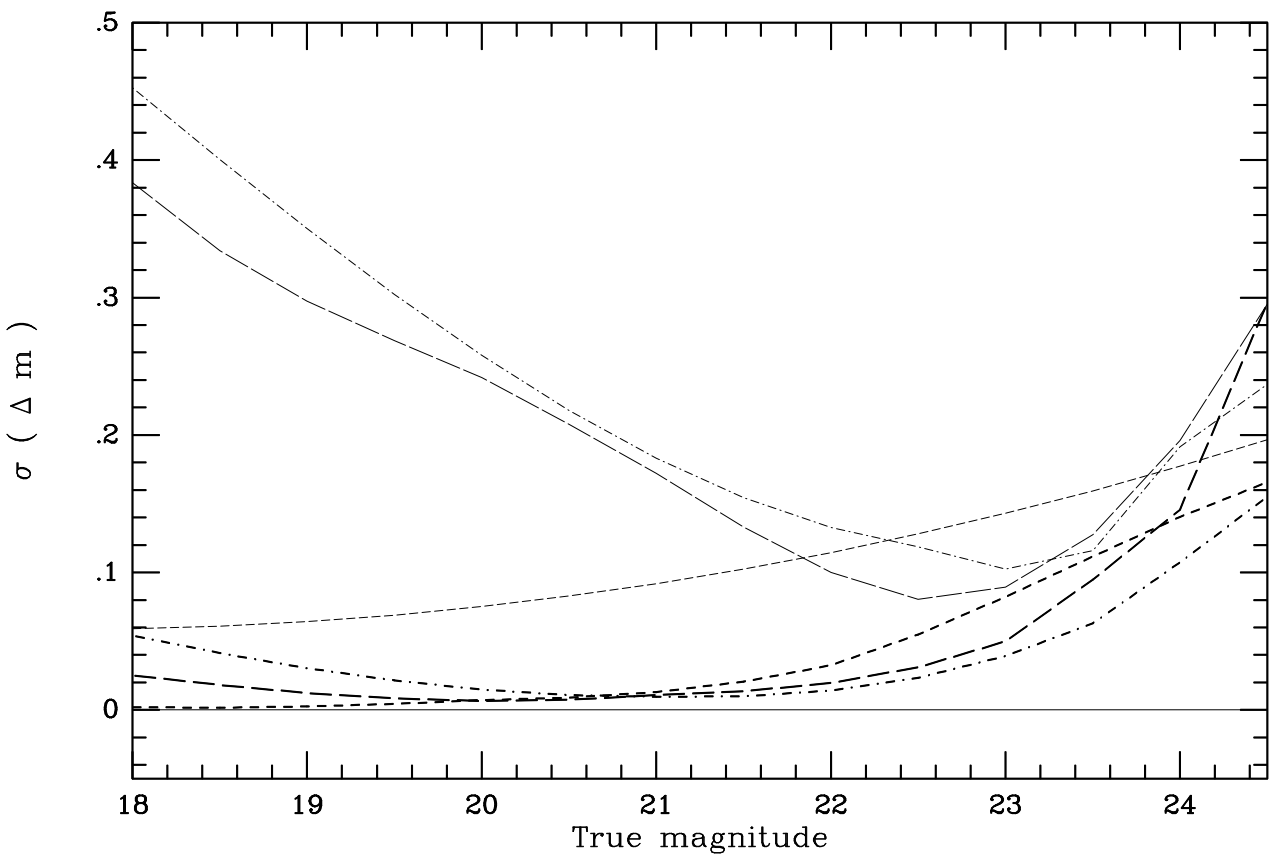

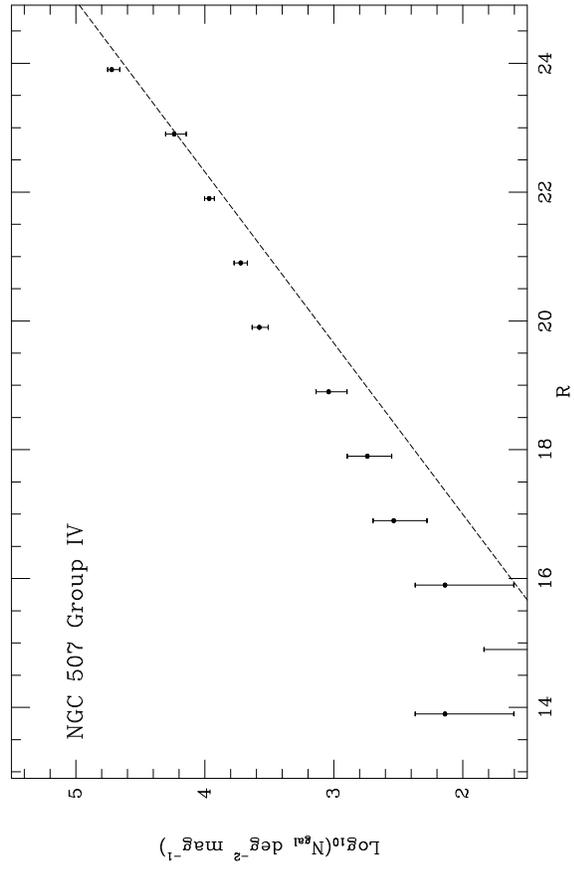

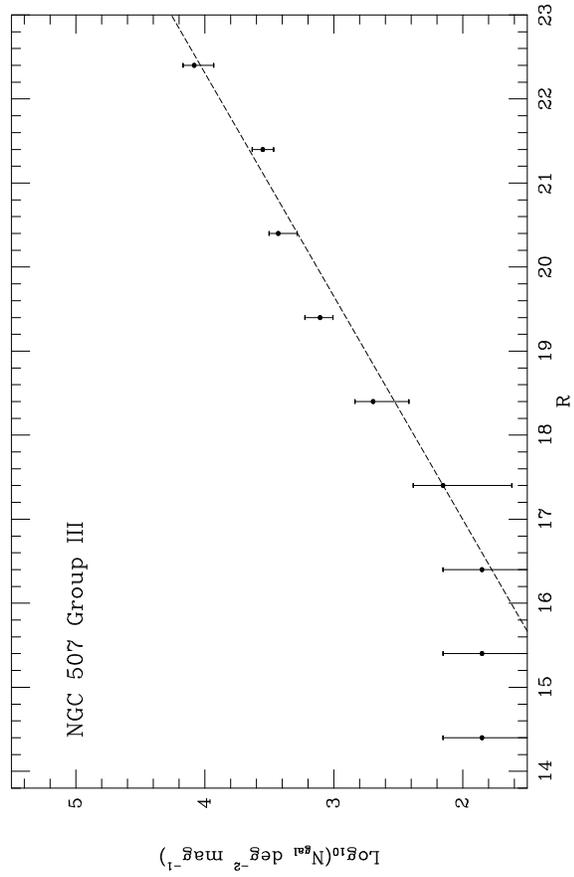

Table 4

Abell 262 Luminosity Function

|  |  | $-15.67 < M_R < -9.67$ | $-15.67 < M_R < -10.67$ |
|---|---|---|---|
| $\eta = 1$ | Fields I + II | $\alpha = -1.84^{+0.28}_{-0.36}$ | $\alpha = -1.93^{+0.34}_{-0.65}$ |
|  | Field I only | $\alpha = -1.90^{+0.22}_{-0.28}$ | $\alpha = -2.08^{+0.32}_{-0.51}$ |
| $\eta = 2$ | Fields I + II | $\alpha = -1.83^{+0.50}_{-1.35}$ | $\alpha = -1.92^{+0.50}_{-\infty}$ |
|  | Field I only | $\alpha = -1.88^{+0.34}_{-0.65}$ | $\alpha = -2.08^{+0.42}_{-1.55}$ |
| $\eta = 3$ | Fields I + II | $\alpha = -1.83^{+1.20}_{-\infty}$ | $\alpha = -1.92^{+1.29}_{-\infty}$ |
|  | Field I only | $\alpha = -1.88^{+0.58}_{-1.57}$ | $\alpha = -2.08^{+0.69}_{-\infty}$ |

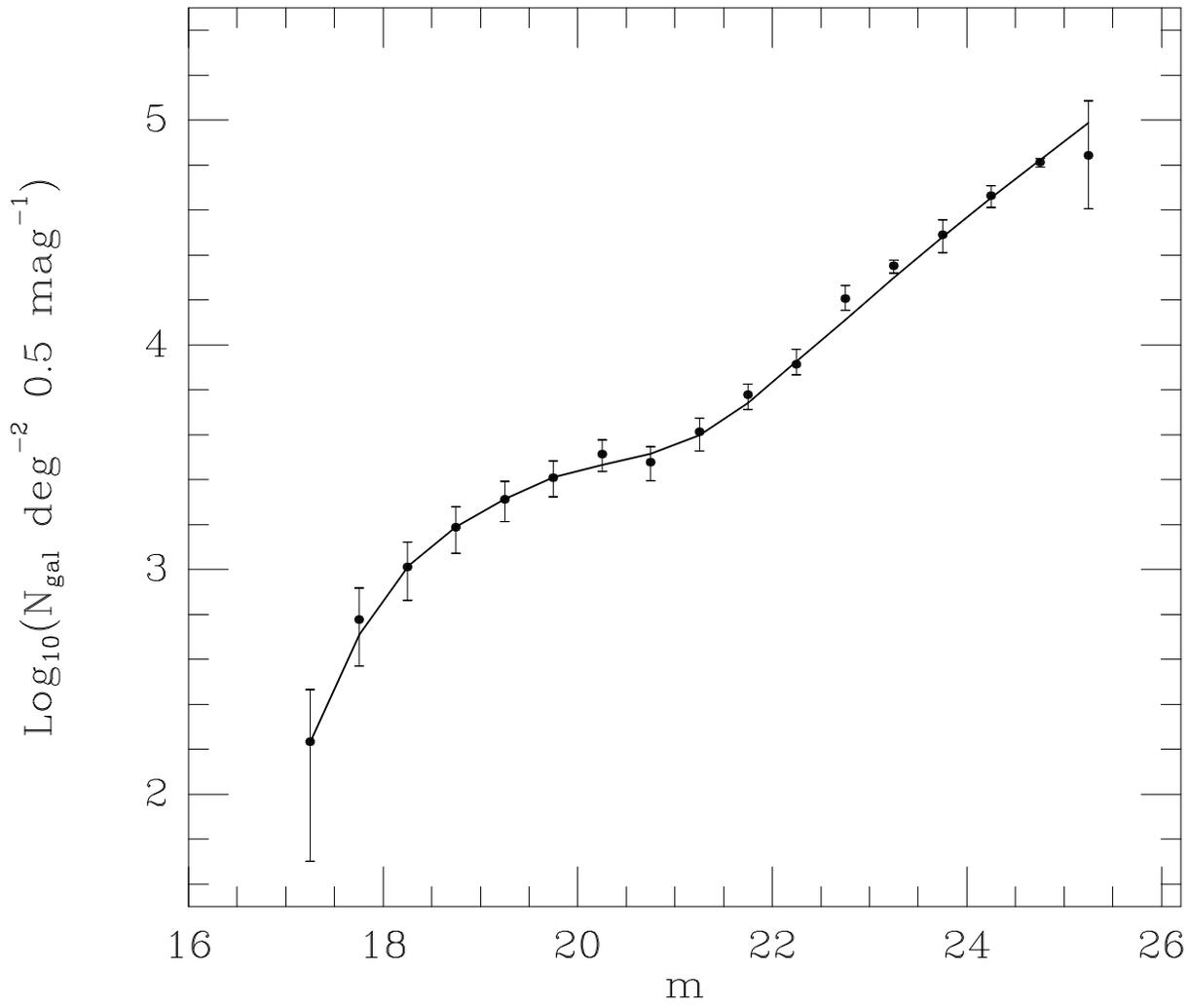

Table 5

Pegasus Luminosity Function

|  | $-15.84 < M_R < -8.84$ | $-15.84 < M_R < -10.84$ (Field I only) |
|---|---|---|
| $\eta = 1$ | $\alpha = -2.50^{+0.33}_{-0.53}$ | $\alpha = -1.93^{+0.63}_{-1.45}$ |
| $\eta = 2$ | $\alpha = -2.41^{+0.51}_{-1.19}$ | $\alpha = -1.91$ [$\alpha$ unconstrained at the $1\sigma$ confidence level] |
| $\eta = 3$ | $\alpha = -2.39^{+0.70}_{-\infty}$ | $\alpha = -1.90$ [$\alpha$ unconstrained at the $1\sigma$ confidence level] |

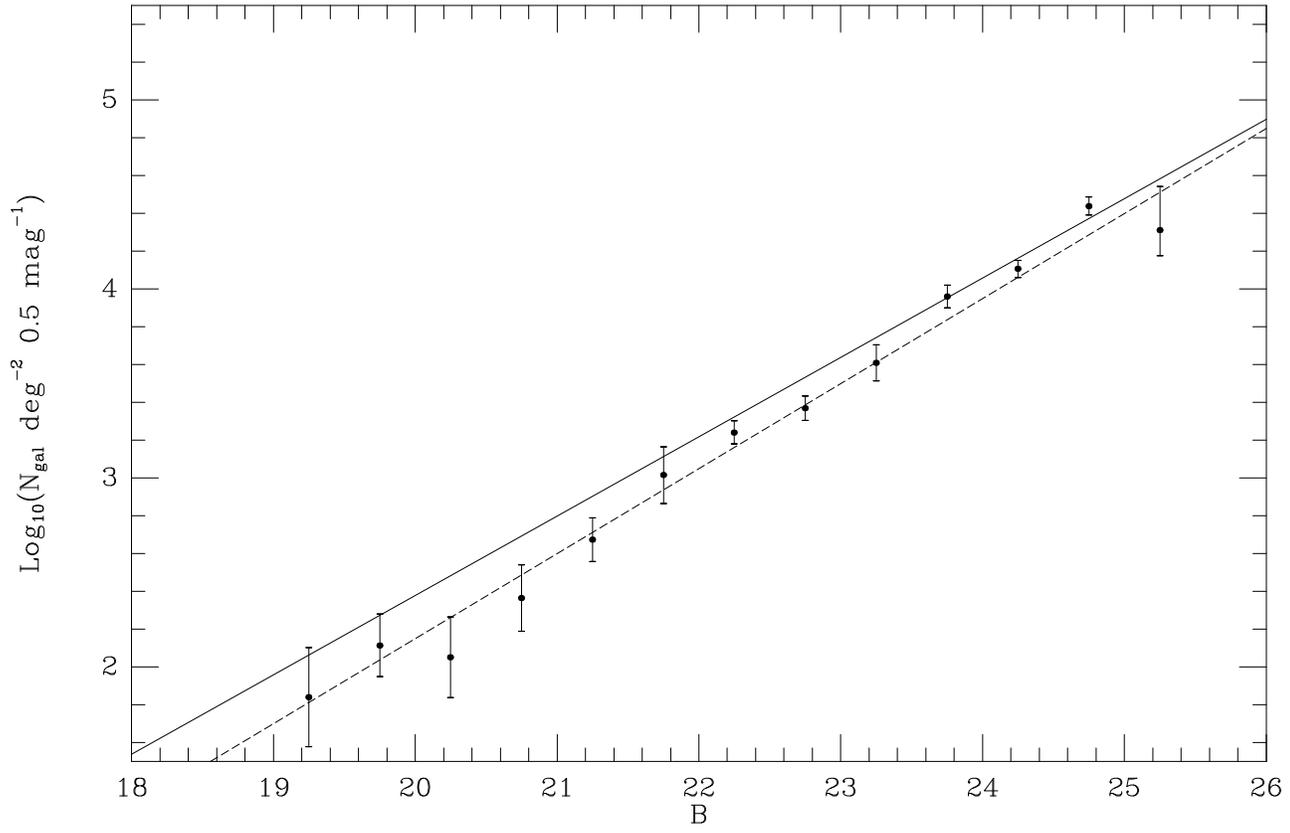
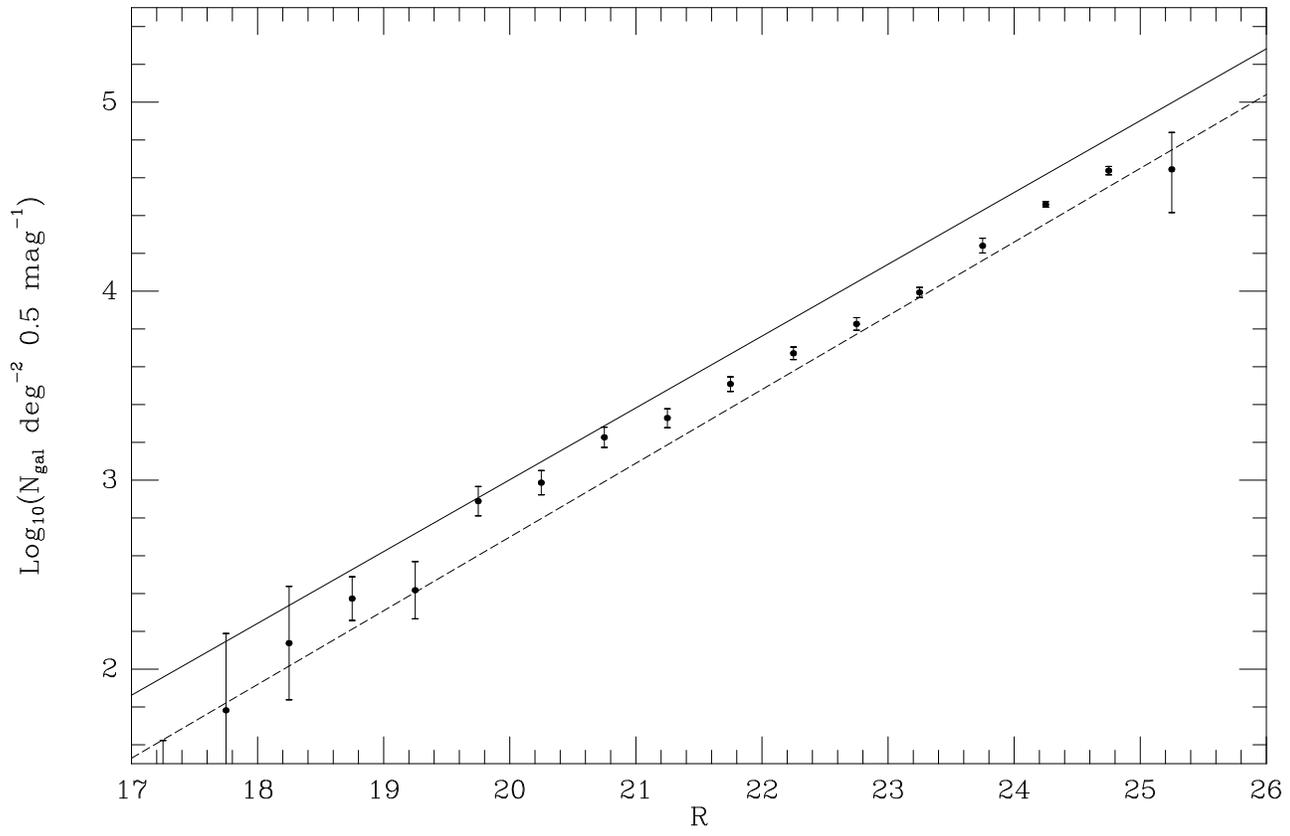

Table 6

NGC 507 Group Luminosity Function

|  | $-15.70 < M_R < -9.70$ | $-15.70 < M_R < -10.70$ |
|---|---|---|
| $\eta = 1$ | $\alpha = -1.62^\dagger$ | $\alpha = -1.42^\dagger$ |
| $\eta = 2$ | $\alpha = -1.64^\dagger$ | $\alpha = -1.43^{+0.40}_{-0.37}$ |
| $\eta = 3$ | $\alpha = -1.67^{+0.49}_{-0.75}$ | $\alpha = -1.46^{+0.68}_{-0.64}$ |

$^\dagger$data is not consistent with a power-law fit at th $1\sigma$ confidence level.

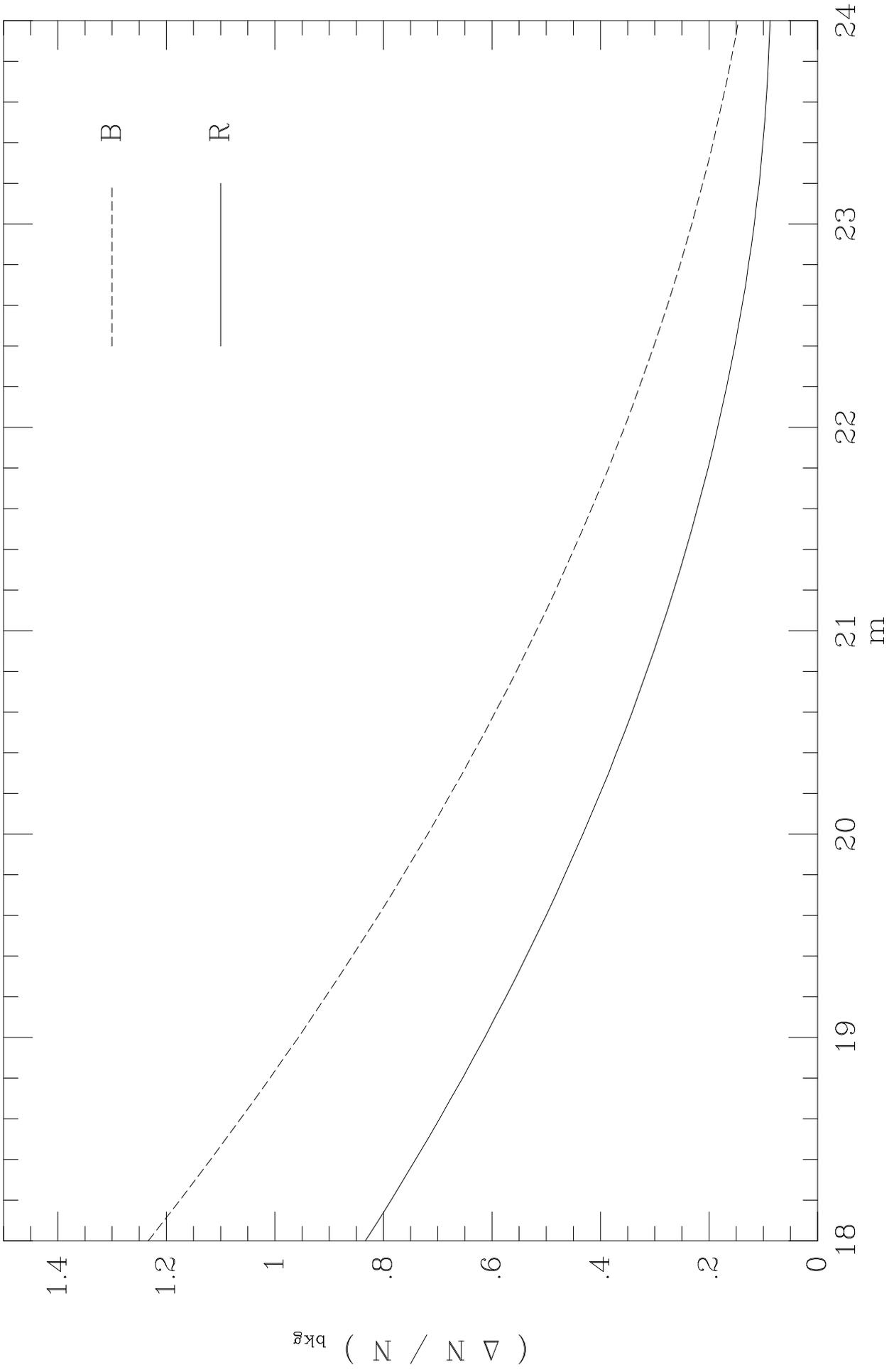

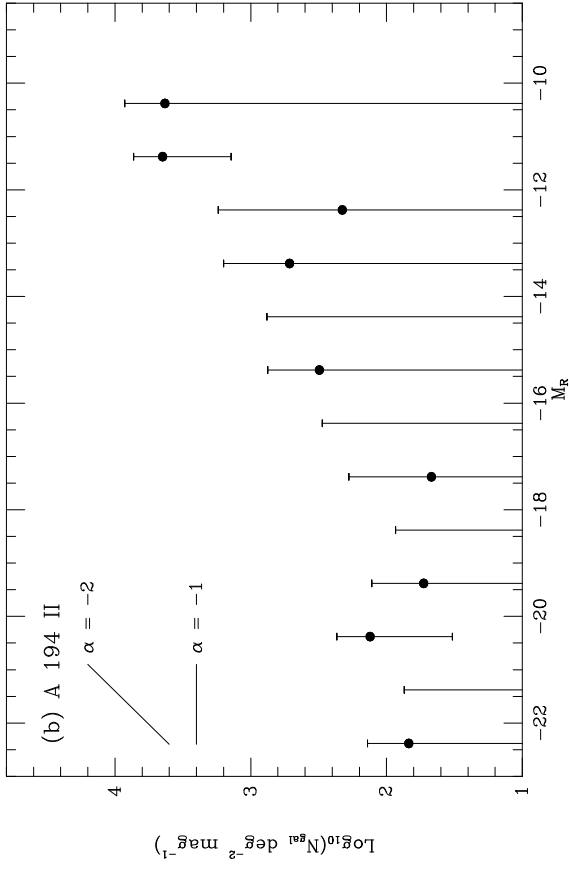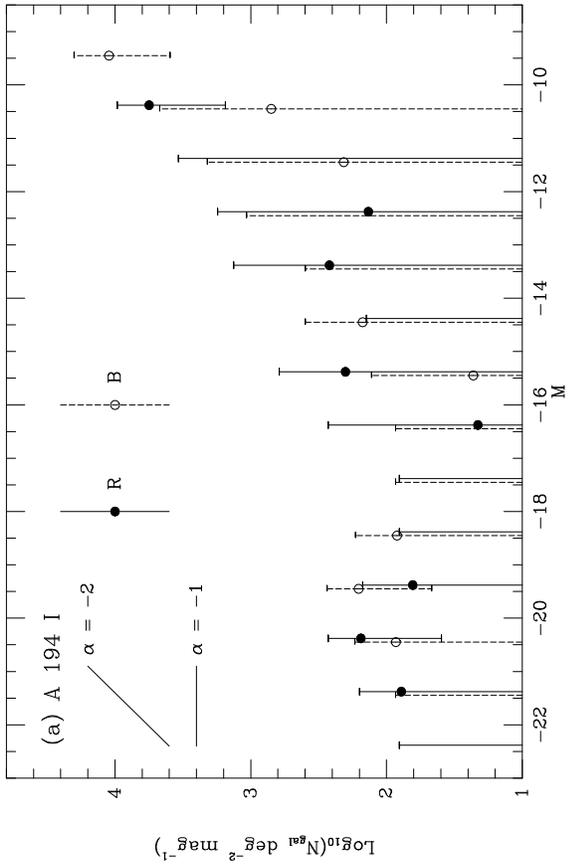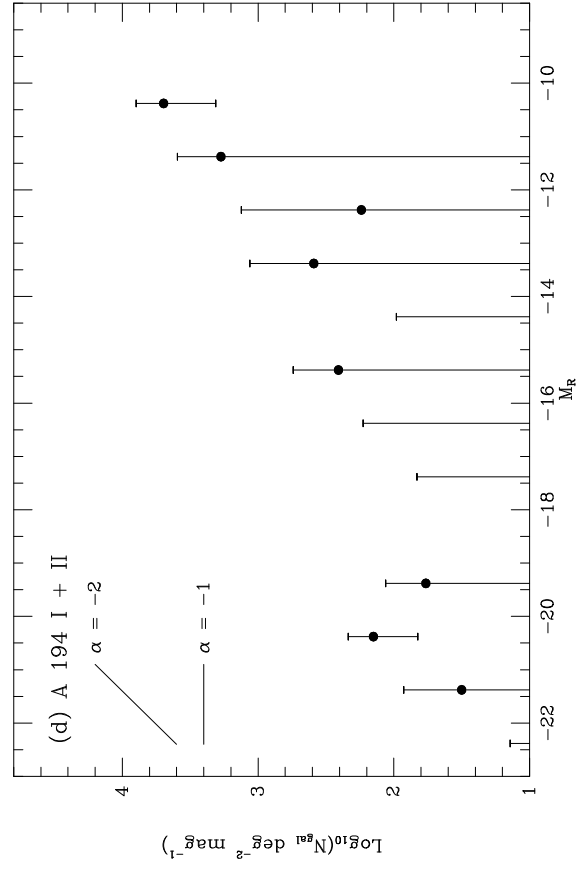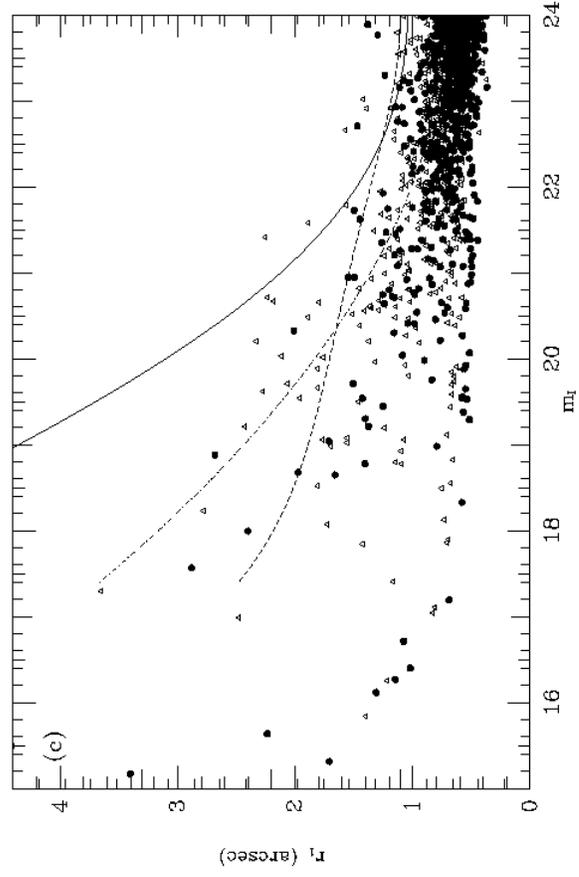

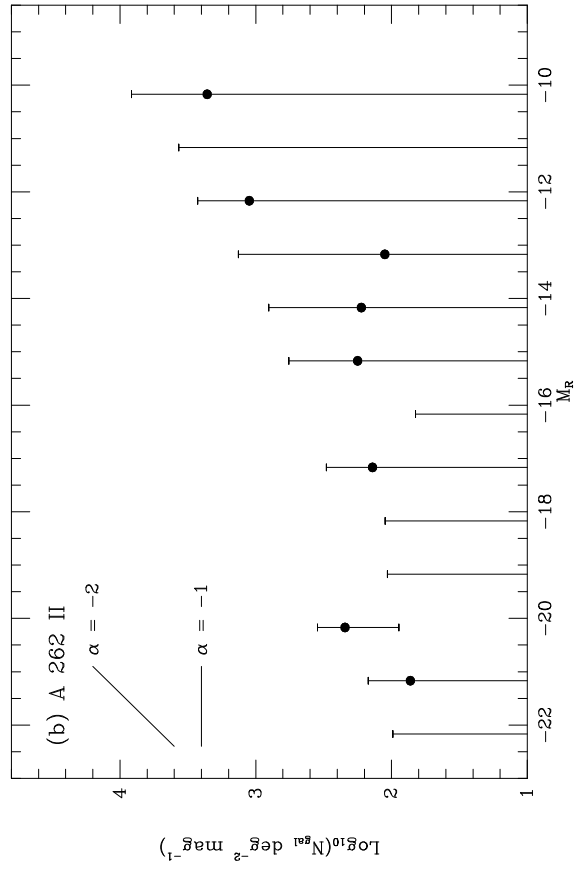
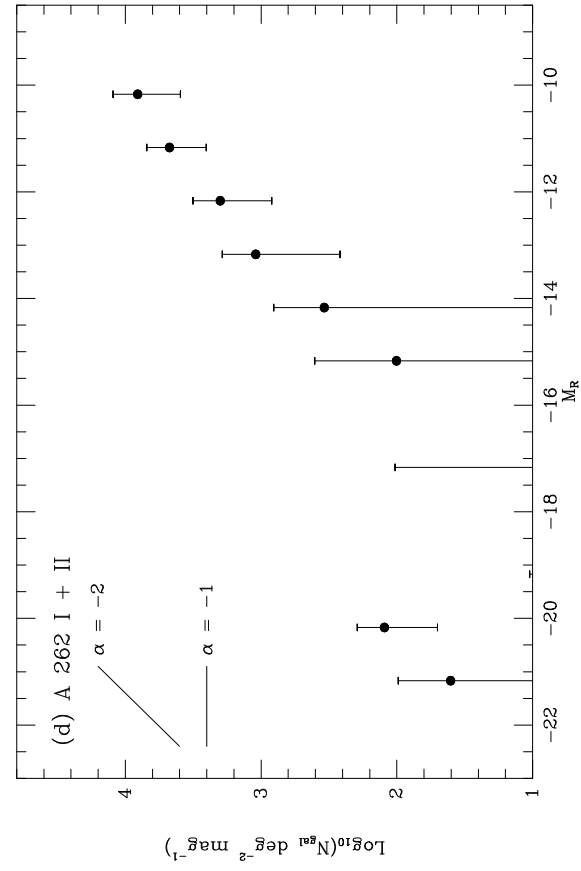
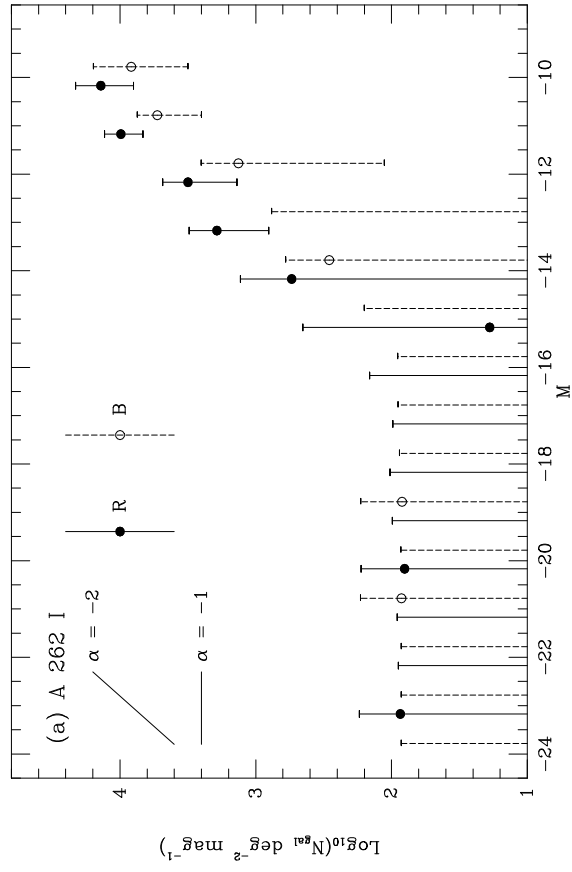
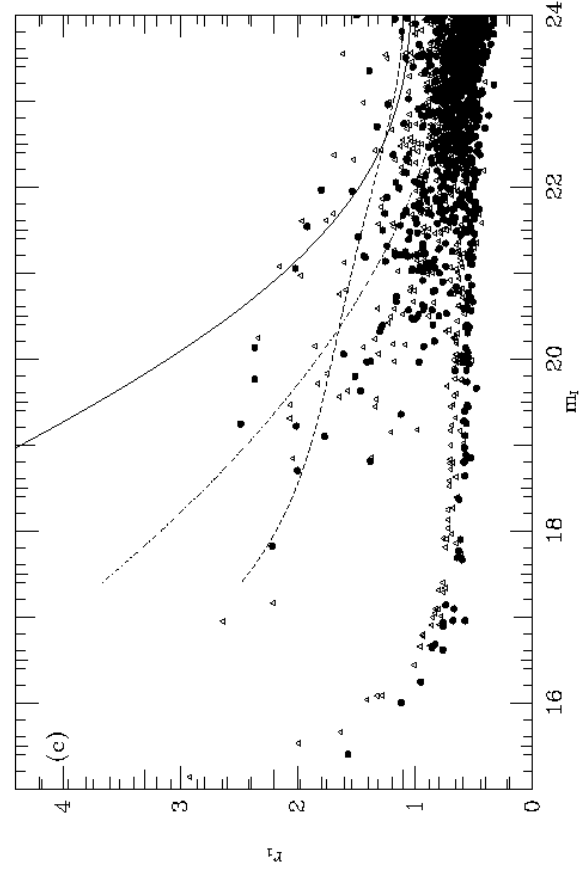

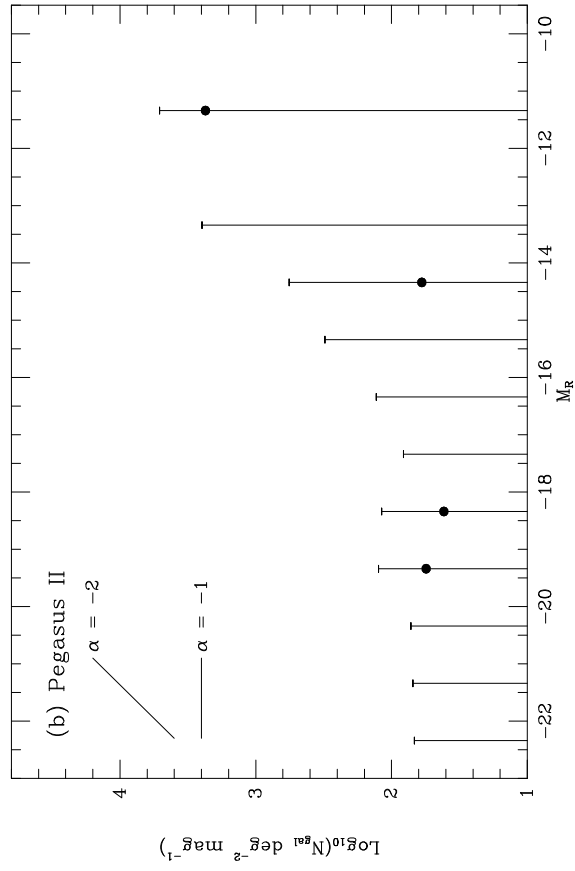
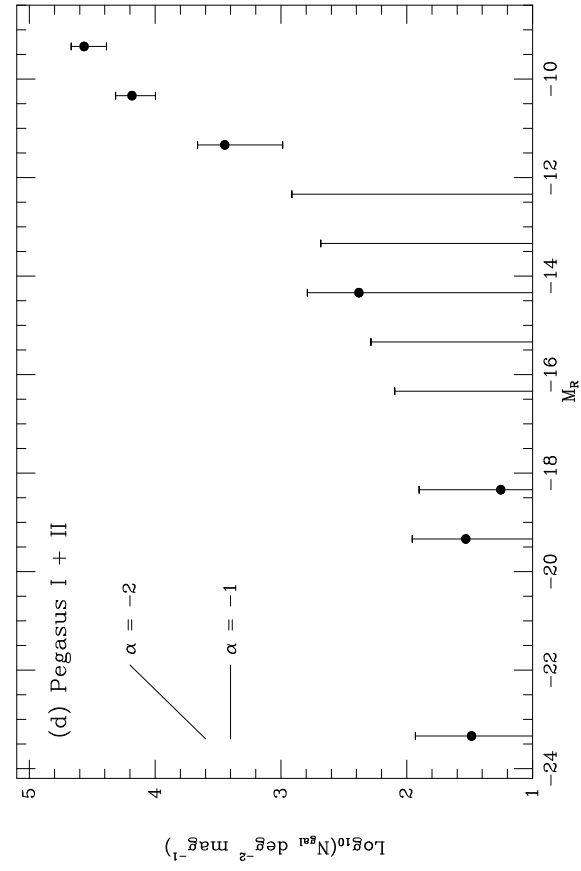
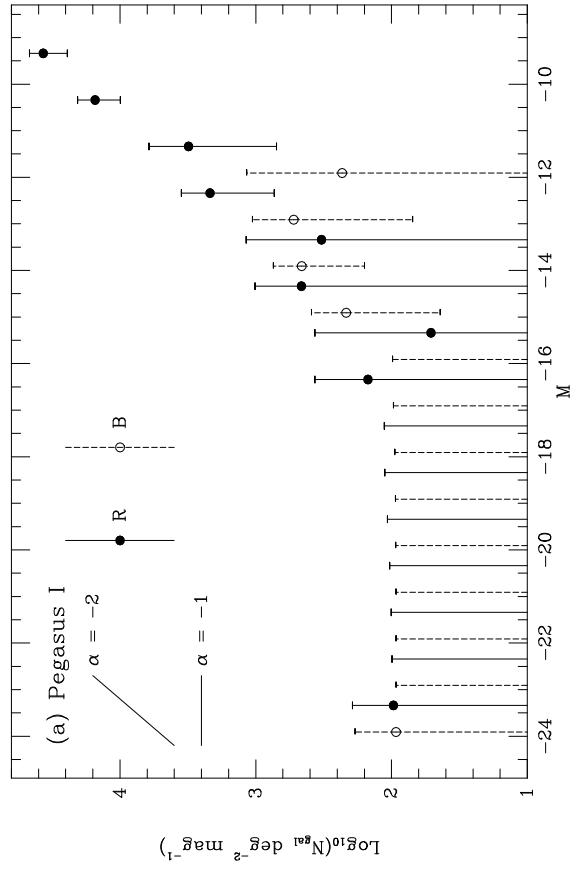
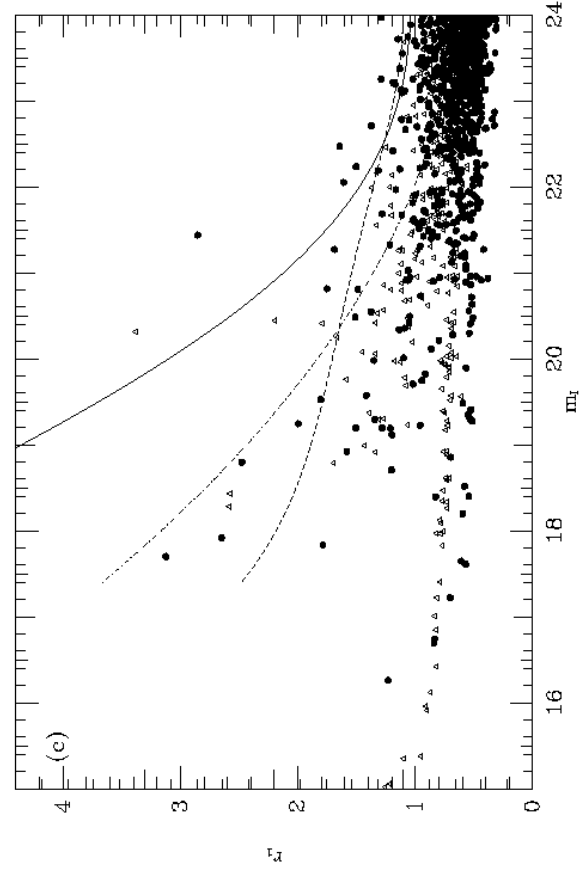

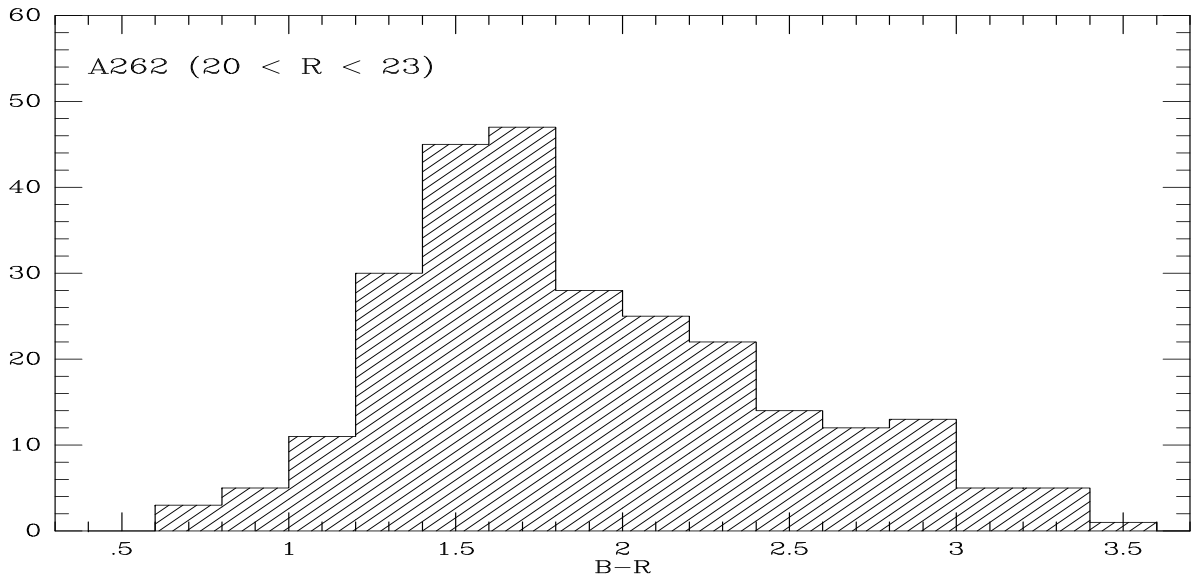
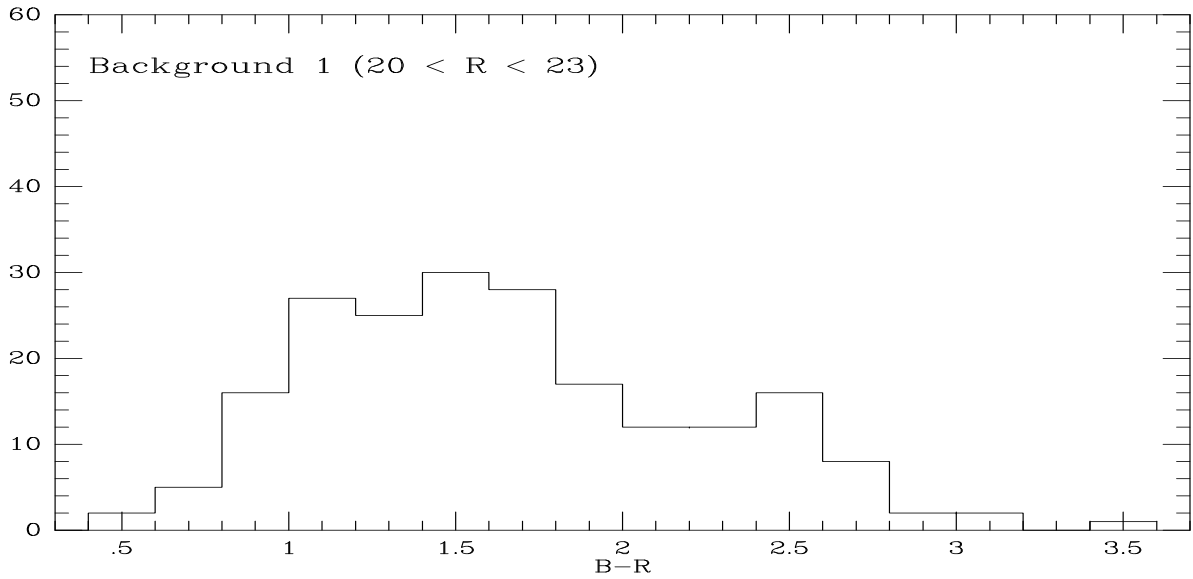
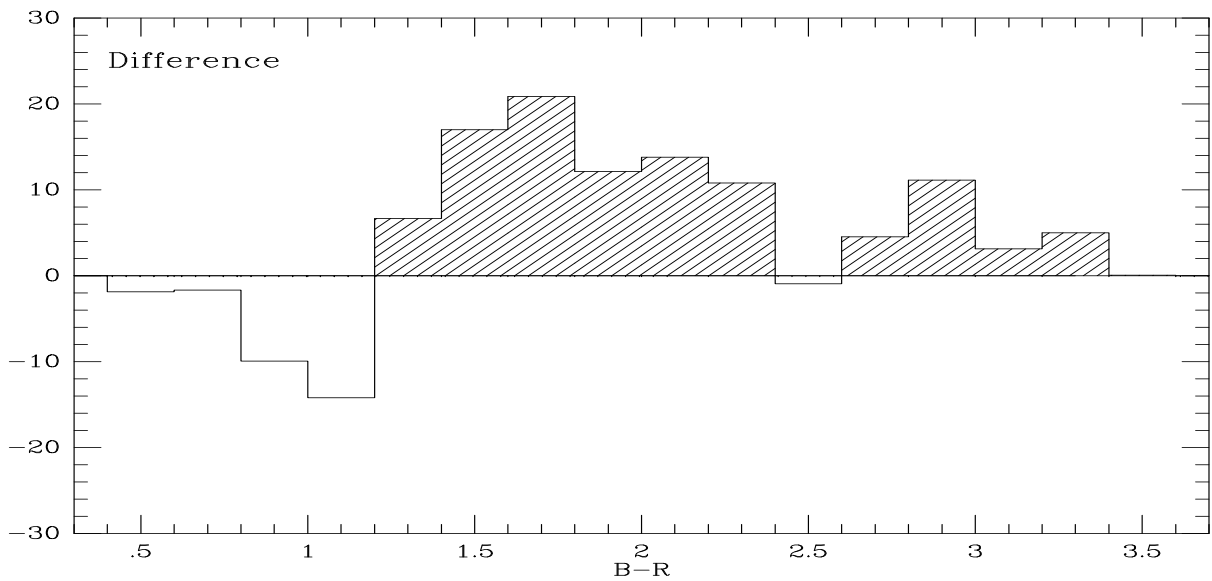

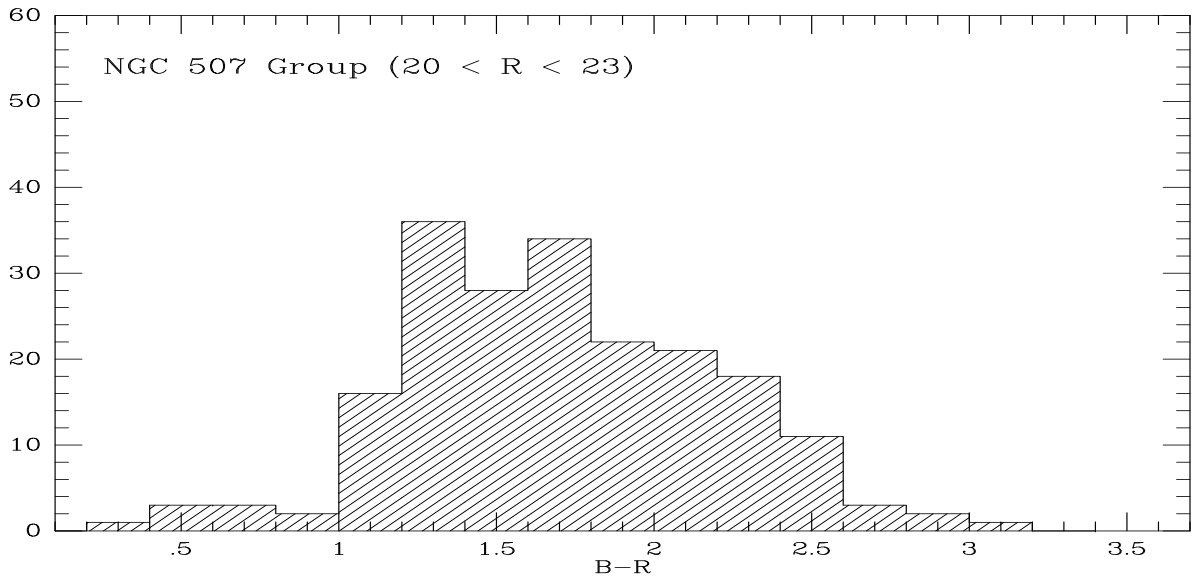
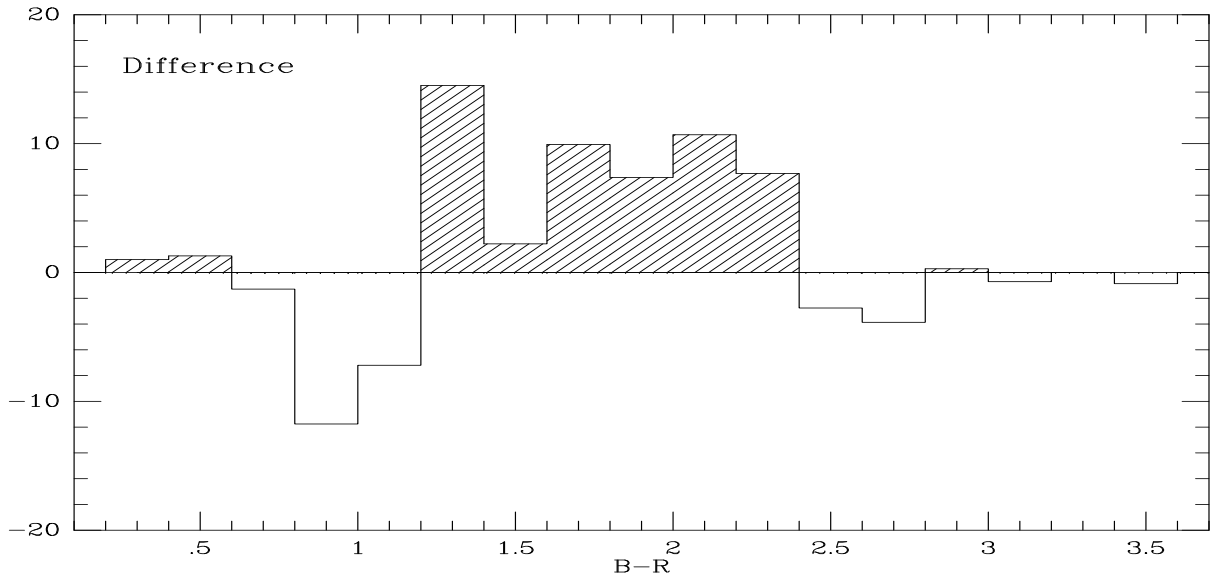

# The luminosity function of dwarf galaxies in four spiral-rich clusters

Neil Trentham

Institute for Astronomy, University of Hawaii

2680 Woodlawn Drive, Honolulu HI 96822, U. S. A.

email : nat@newton.ifa.hawaii.edu

# ABSTRACT

We measure luminosity functions in the cores of four spiral-rich, poor clusters of galaxies at median redshift $z = 0.016$. In the red magnitude range $-14 < M_R < -10$, our data imply that the luminosity functions $\phi(L) \propto L^\alpha$ are steep, $-1.8 < \alpha < -1.6$, in the central 200−300 kpc of Abell 262 and of the NGC 507 Group. Abell 194 also shows signs of a steep luminosity function, $\alpha < -1.6$, in this magnitude range. In Pegasus, the dwarf galaxy density is too low relative to the background to let us constrain $\alpha$.

The NGC 507 Group and Abell 194 have been interpreted as clusters that are forming today, based on morphology and velocity structure. The high spiral galaxy fraction in Abell 262 relative to clusters like Virgo and Coma also suggests that it is young. We therefore suggest that steep luminosity functions in the range $-14 < M_R < -10$ may be a universal feature of young clusters and possibly of the field. If this is true, then the observed paucity of gas-rich galaxies in such environments suggests that we are finding galaxies similar to the low-surface-brightness, dark-matter-dominated dwarf spheroidal galaxies seen locally and in Virgo. This interpretation is also consistent with the distribution of colors and sizes of the faint galaxies in Abell 262. If we are indeed detecting dwarf spheroidal galaxies and if they are as numerous relative to bright galaxies in the field as they are in the young clusters observed here, then the contribution of their halos to the cosmological mass density is $\Omega_{\text{dSph halo}} \approx 0.01$. This is much smaller than values of $\Omega$ derived from dynamical measurements.

**Key words:** galaxies: clusters: luminosity function − galaxies: clusters: individual: A194, A262, Pegasus, NGC 507 Group − galaxies: photometry

# 1 INTRODUCTION

Dwarf galaxies provide valuable insight into galaxy formation processes. This is not surprising. Most evidence (e.g., Blumenthal et al. 1984; Bardeen et al. 1986) suggests that galaxies formed from a post-recombination fluctuation spectrum $\delta(k)$ whose power spectrum $|\delta(k)|^2 \propto k^n$, $k = $ wavenumber, has index $n \approx -2$. This value of $n$ results in a mass distribution in which small galaxies are much more common than large ones (White & Rees 1978). Therefore low-luminosity galaxies are expected to be a substantial constituent of the universe, so their statistical properties are an important probe of galaxy formation theories.

The most directly observable statistical property of dwarf galaxies is the luminosity function $\phi(L)$, defined as the number density of galaxies per unit luminosity $L$. The dwarf galaxy luminosity function is predicted naturally by most theoretical models of galaxy formation. It depends mainly on the mechanisms that control star-formation efficiencies in low-mass galaxies; these differ significantly between published models.

Many studies of $\phi(L)$ have concentrated on clusters of galaxies. Then the distances to the lowest luminosity galaxies are known, at least statistically, so the luminosity function can be derived from photometry alone. This in turn means that $\phi(L)$ can be determined to very low luminosities. However, the interpretation of the luminosity function in most clusters is complicated by environmental effects such as the tidal destruction of low-mass galaxies.

In this study, we present luminosity functions for four clusters of galaxies that are likely to be young and dynamically unevolved. They are rich enough so that $\phi(L)$ does not (except in one case) suffer from poor statistics after background subtraction, and they are unevolved enough so that effects like tidal destruction may still be small. This means that a comparison between our results and galaxy formation models can provide constraints on the efficiency of star formation in low-mass galaxies.

Lower luminosity galaxies are increasingly dark-matter dominated. The faintest dwarfs known (Draco and Ursa Minor, $M_B \sim -8$) are $\sim 99$ per cent dark matter by mass – their stellar populations that are not self-gravitating (Aaronson & Olszewski 1987; Olszewski et al. 1996). Kormendy (1988, 1990) has determined the scaling laws that describe how dark matter halo central densities and core radii vary with galaxy luminosity. These laws are derived for nearby late-type and dwarf spheroidal galaxies. We assume that the clusters studied here have dwarf galaxy populations that are similar to the field population; we then combine our luminosity functions with the above scaling laws to estimate the total cosmological density contributed by the halos of low-mass galaxies.

This paper is organized as follows. In Section 2, we review the observations of the luminosity functions of dwarf galaxies in clusters and outline the limitations of previous work. In particular, we explain why so little is known about the faint-end slope $\alpha$. In Section 3, we present our cluster sample. In Section 4, we describe the observations and data analysis. This section includes a description of how we perform the photometry and compute the luminosity functions, emphasizing why we use a surface-brightness–dependant method of making isophotal corrections. It also presents our background number counts and compares them with published results. Section 5 presents the luminosity functions. Finally, in Section 6, we discuss the results in light of recent observational and theoretical developments.

Throughout this paper, we assume that the Hubble constant is $H_0 = 75$ km s$^{-1}$ Mpc$^{-1}$ and that the cosmological density parameter is $\Omega_0 = 1$.

## 2 REVIEW

In this section, we review previous observations of dwarf galaxies in clusters and outline our current understanding of the faint end of the luminosity function.

One of the pioneering studies in this area was Sandage et al. (1985, hereafter SBT85). They measured the Virgo Cluster luminosity function down to $M_B \sim -12$ and found that the faint-end slope was $\alpha = -1.35$. This was steeper (i.e. $\alpha$ is more negative, so that fainter galaxies are more numerous relative to bright ones) than anywhere else known at the time. The Virgo Cluster has therefore become regarded as the prototypical dwarf-rich environment.

SBT85 also showed that the luminosity function at $M_B > -16$ is strongly dominated by dwarf spheroidal (dSph) galaxies. These dSphs (which are called dwarf ellipticals by SBT85, see Binggeli 1994; Kormendy & Bender 1994) are physically different from elliptical galaxies as shown by the observation that they have different scaling laws in the fundamental plane parameter correlation diagrams (Kormendy 1985, 1987b; Binggeli & Cameron 1991; Ferguson & Binggeli 1994). Unlike ellipticals, dSphs have lower surface brightnesses as $L$ decreases (for example, in the Local Group, Draco and Ursa Minor have far lower surface brightnesses than NGC 185 and NGC 205). They are also increasingly dark-matter dominated at lower luminosities (Kormendy 1990). The faintest dSph galaxies (Draco and Ursa Minor, $M_B \sim -8$) are $\sim 99$ per cent dark matter by mass and so have stellar populations that are not self-gravitating (e.g., Aaronson 1983, Aaronson & Olszewski 1987, Kormendy 1987a, 1990; Lake 1990; Pryor & Kormendy 1990; Olszewski et al. 1996). The most straightforward explanation is that lower mass galaxies are more susceptible to baryonic blowout by supernova-driven galactic winds during an early epoch of star formation (Saito 1979; Dekel & Silk 1986). Their precursors were then gas-rich star-forming dwarfs presumably similar to the gas-rich dwarf irregular (dIrr) galaxies that we see locally (Kormendy & Bender 1994). The similar positions of dSph and dIrr galaxies in the fundamental plane correlation diagrams (Kormendy 1985, 1987b) are also suggestive of an evolutionary link.

In a subsequent study of the Virgo Cluster, Impey et al. (1988, hereafter IBM88) discovered a number of low-surface-brightness (LSB) galaxies that SBT85 missed because their surface-brightness detection threshold was too high. Impey and collaborators suggested that the SBT85 sample is heavily incomplete at the faintest magnitudes and calculated completeness corrections to $\phi(L)$ that took the newly-discovered LSB galaxies into account. These corrections suggested that the luminosity function might be as steep as $\alpha = -1.7$. If the luminosity function is this steep, then small, dark-matter-dominated dSph galaxies might contribute significantly to the cluster mass. However, the results of IBM88 are model-dependant. Given the large difference between the IBM88 and SBT85 results, the faint-end slope of the Virgo Cluster luminosity function remains uncertain.

For richer and more distant clusters of galaxies (like Coma), the dwarf galaxy luminosity function is even more uncertain. Schechter (1976) showed that the luminosity function of galaxies in clusters is well fit by the function

$$\phi(L) = \frac{\phi^*}{L^*} \left(\frac{L}{L^*}\right)^{\alpha^*} \exp\left(-\frac{L}{L^*}\right). \tag{1}$$

Here $\phi^*$ is a normalization density, and the shape of the luminosity function is described by the parameters $L^*$, a cutoff luminosity above which galaxies are rare, and $\alpha^*$, the faint-end slope. Most studies of the luminosity functions of rich clusters present the results as Schechter function fits (e.g. Lugger 1986; Oegerle & Hoessel 1989; see Binggeli et al. 1988 for a review). But these studies typically reach only to $M_B \simeq -18$. Dwarf spheroidal galaxies, as defined by their position in the fundamental plane (Kormendy 1987b, Ferguson & Binggeli 1994), are a trace population brighter than $M_B = -18$; they do not dominate the number counts until $M_B > -16$ (Binggeli 1987; Binggeli et al. 1988). But Binggeli and collaborators show that only dSph galaxies have luminosity functions $\phi \propto L^{\alpha^*}$; other types of galaxies – i.e., the ones to which most published studies are sensitive – have bounded luminosity functions. So almost no published measurements provide information about $\alpha^*$. Instead, the derived values of $\alpha^*$ are determined mainly by the strong parameter coupling with $L^*$. For example, Lugger (1986) shows that excluding a single bright galaxy from a Schechter function fit to Abell 779 increases $\alpha^*$ by 0.7!

The Schechter function was proposed to fit the total luminosity functions of all types of galaxies. But a total luminosity function is really the sum of the differently-shaped luminosity functions of different types of galaxies (Binggeli 1987; Binggeli et al. 1988). Also, since different types of galaxies reside in different parts of parameter space, it is very likely that different physical processes are at work in producing their luminosity. For these reasons, it has become customary to not attribute physical significance to Schechter function fits. Even if we were to measure the luminosity function faint enough so that $\alpha^*$ is free of parameter coupling errors, it would still tell us more about the dSph-to-other galaxy ratio (i.e., about the relative numbers of galaxies in different parts of the fundamental plane correlations) than about the faint-end slope of the dwarf galaxy luminosity function. Making double Schechter function fits is not the answer: the uncertainties are huge because of parameter coupling. In this work, we will not use Schechter fits to measure $\alpha^*$. Rather, we will determine $\alpha^*$ directly by fitting power laws to the luminosity functions. In this paper we use $\alpha$ to describe the logarithmic slope of the luminosity function, and $\alpha^*$ to describe the best-fitting Schechter parameter.

Historically, studies of cluster luminosity functions much fainter than $M_B \approx -18$ were difficult because of contamination from background galaxies. However, large CCDs now allow us to determine background counts and their variance accurately (Tyson 1988, Lilly et al. 1991, Bernstein et al. 1995). As a result, we can make background subtractions to faint magnitudes with increased confidence. This has permitted various authors to make direct measurements of the dwarf galaxy luminosity function in clusters. Bernstein et al. (1995, hereafter B95) attempt to measure the Coma cluster dwarf galaxy luminosity function to $M_R = -9.4$. However their results suffer from severe globular cluster contamination in the halo of NGC 4874. As Figure 5 of their paper shows, they are unable to constrain the galaxy luminosity function much fainter than $M_R = -14$. Thompson & Gregory (1993) survey the Coma cluster over a much larger area using photographic data. They determine the luminosity function down to $M_B = -15$ ($M_R \sim -16$) and find $\alpha^* \sim -1.4$ at the faint end of their sample. Driver et al. (1994b) measure the luminosity function of the extremely rich cluster Abell 963 at $z = 0.2$ down to $M_R = -14$ and find a slope of $\alpha^* = -1.8$. An even steeper dwarf galaxy luminosity function ($\alpha \approx -2.2$) has since been claimed for the cD halo of Abell 2199 at $z = 0.029$ (De Propris et al. 1995). However, in a complementary study (Trentham & Kormendy 1996), we argue that the luminosity function brighter than $M_R = -14$ is flat in such environments, and we discuss the reasons why the above authors found results so different from this.

In any case, clusters like Virgo and Coma are not typical environments for galaxies. Their crossing times are less than a Hubble time, and their smooth X-ray halos and large elliptical galaxy fractions suggest that the galaxies in them have evolved via cluster-related processes. Therefore results like those quoted above do not necessarily tell us anything directly about the cosmological distribution of dSphs and the type of dark matter that is in their halos. Poor, spiral-rich groups and clusters represent an environment that is far more typical cosmologically. Many of the giant galaxies in the universe reside in such environments. Tully (1988) has shown that the luminosity function of gas-rich dwarf galaxies in such environments is not steep ($\alpha \simeq -1$). Comparison between the results of Tully (1988) and those of Ferguson & Sandage (1991) then suggests that diffuse groups may have a substantial population of low-luminosity gas-poor galaxies at $M_B \sim -15$. These would presumably be dSph galaxies.

The aim of the present study is to look for such a population of galaxies by measuring the luminosity function down to faint limits in spiral-rich clusters. The Local Group is such an environment, but we cannot measure its luminosity function because of small number statistics and incompleteness (galaxies with $M_V \simeq -13$ are still being discovered: Ibata et al. 1994). Instead, we shall measure the luminosity functions of four poor clusters of galaxies. All show signatures of being young and dynamically unevolved, but all are rich enough to give us a good chance to measure $\alpha$ after background subtraction. Assuming that these poor clusters are sufficiently unevolved that they are representative of the universe, we then combine the luminosity functions with the dark-matter scaling laws for late-type and dSph galaxies to estimate the cosmological density of the dark matter that is in dSph halos. The scaling laws come from studies of local group dSphs and nearby late-type galaxies and so can be applied in this context with reasonable confidence. We shall also compare our results with detailed theoretical models of the galaxy luminosity function implied by various galaxy formation scenarios (e.g., Kauffmann et al. 1993; White & Kauffmann 1994; Babul & Ferguson 1996).

This work is complementary to studies of the faint end of the field galaxy luminosity function. The deepest work on this subject (Marzke et al. 1994) suggests that $\alpha^* \approx -1.8$ for dIrr galaxies at faint magnitudes; this would be inconsistent with Tully's (1988) result for diffuse nearby groups if these groups are representative of the field. However, the statistics of the field sample are poor at faint magnitudes. By determining the luminosity function from photometry alone, we can probe to fainter magnitudes than the field studies ($M_B \sim -9$ as compared to $M_B \sim -14$). What we give up is some confidence that the environments we study are cosmologically representative. Also, our clusters are richer than the nearby groups studied by Ferguson & Sandage (1991). However, we can measure the luminosity function deeper than they did, because the galaxy densities in our clusters are higher.

## 3 SAMPLE

Our cluster sample is given in Table 1. For each cluster, we list the Abell (1958) richness, redshift $z$, central velocity dispersion $\sigma$, X-ray luminosity $L_x$, $B$-band Galactic extinction $A_B$, and the critical surface density $\Sigma_c$ for gravitational lensing of distant background sources (see below).

Abell 194 (hereafter A194) is a prototypical "linear" cluster (Chapman et al. 1988), which suggests that it has not undergone significant gravitational relaxation and so may reflect the conditions under which clusters formed. This is further implied by its large spiral fraction – this **does not** follow Dressler's (1980) morphology-density relation (there is no evidence for morphological segregation, Chapman et al. 1988). Abell 262 (hereafter A262) is similar in richness and X-ray properties to Virgo but has a much higher spiral fraction (Fanti et al. 1982). The sequence Fornax−Virgo−A262 is one of increasing spiral fraction; if it is also an evolutionary sequence, then the variation in the dwarf galaxy luminosity function along this sequence will be a useful probe of cluster evolution theories (see SBT85 and Ferguson

& Sandage 1988 for the Virgo and Fornax cluster luminosity functions, respectively). Still, Abell 262 is the most evolved cluster in our sample; it is therefore least representative of the field. Pegasus has an excess of star-forming, early-type galaxies. Vigroux et al. (1989) suggested that it is a low-$z$ analogue of the Butcher-Oemler (1978) clusters. If the infall of galaxies into groups and clusters is part of the explanation of the Butcher-Oemler effect (Kauffmann 1995), then Pegasus may be a local example of a cluster in formation. Finally, the NGC 507 Group is a group of gas-rich spiral galaxies lying on the filamentary main ridge of the Pisces-Perseus Supercluster. It was interpreted as a cluster in formation by Sakai et al. (1994), who studied its distribution of galaxies in position and velocity space.

## 4 OBSERVATIONS AND PHOTOMETRY

In this section, we describe our data collection, reduction, and analysis. Section 4.1 describes the observations. In Section 4.2, we describe in detail our galaxy detection strategy and photometry techniques, and we explain our methods of making isophotal magnitude corrections. In Section 4.3, we describe how we apply these methods to our data and we present plots of number counts *versus* magnitudes for each of our cluster and background fields. In Section 4.4, we present the results for our background fields and discuss the statistics of the number counts. Section 4.5 gives the results for our cluster fields and describes how we make background subtractions to convert the number counts into luminosity functions. The luminosity functions are then discussed individually in Section 5.

### 4.1 Observations and CCD preprocessing

Table 2 gives our observing log. All observations were made at the f/10 Cassegrain focus of the University of Hawaii 2.2 m telescope on Mauna Kea, except for the observations of Background 3, which were taken at the f/8 Cassegrain focus of the Canada-France-Hawaii 3.6 m Telescope (CFHT) with the MOS-SIS spectrograph in direct imaging mode. The detector was a thinned Tektronix 2048 × 2048 CCD for the 2.2 m observations (scale $0''\!.22$ pixel$^{-1}$, field of view $7'\!.5 \times 7'\!.5$) and an unthinned LORAL 2048 × 2048 CCD (scale $0''\!.31$ pixel$^{-1}$, field of view $10.\!'1 \times 10.\!'1$) for the CFHT observations. In Table 2 we also present, for each of the fields we surveyed, the Mould filters used, the total field size surveyed (this is less than the CCD field size because we dithered our images by a few arcseconds in order to reject bad pixels), the field coordinates, the total exposure time (typically obtained in 10 minute integrations), the mean extinction $<X>$ from clouds, and the seeing. When conditions were not photometric but the cloud cover was thin, we took exposures and later calibrated them using short exposures of the same field taken in photometric conditions.

Images were bias-subtracted and flat-fielded using twilight flats. They were then registered and combined, taking care to reject bad pixels and those affected by cosmic rays (these were identified by their appearance in only one out of several dithered images). Images taken under non-photometric conditions were weighted accordingly during this calculation. Exposures taken when the seeing was substantially worse than one arcsecond were discarded; we shall see in the following sections that the transformation between isophotal and true magnitudes can have huge uncertainties when the seeing is poor. The photometric system used in this paper is Landolt's Johnson (UBV) – Cousins (RI) magnitude system. Calibration was provided by observations of $\sim 30$ Landolt (1992) standard stars per night; the extinction corrections and magnitude transformations are accurate to about 2 per cent. Because we need to make accurate background subtractions and because the background number counts are a steep function of magnitude, accurate photometry is important.

The final $R$-band images are displayed in Figure 1. They show a range of $R \sim 22$ galaxy densities. Of the cluster fields, frame (l) is the richest, and frame (d) is the poorest.

### 4.2 Photometry: Surface-brightness-dependant isophotal corrections

### 4.2.1 The method

To compute the luminosity function, we must determine for each galaxy the total apparent magnitude $m^T = M^T + \mu$ and its uncertainty, where $M^T$ is the absolute magnitude and $\mu$ is the distance modulus. Here we use capital-letter subscripts to denote measured quantities and capital-letter superscripts to denote intrinsic quantites. What we measure is an isophotal magnitude at detection isophote $m_I$ and a light distribution $I(\mathbf{r})$. Here $\mathbf{r}$ represents a position vector in the plane of the sky. The transformation of $m_I$ and $I(\mathbf{r})$ to $m^T$ depends on many things, some of which we measure (like the rms sky noise $\sigma_{\rm rms}$ and the seeing $b_{\rm FWHM}$) and others of which are unknown (like the intrinsic projected brightness profile $\rho(\mathbf{r})$).

Our approach is as follows. We assume that detected galaxies have exponential projected brightness profiles; this is a good approximation both for cluster dSphs and dIrrs (Faber & Lin 1983, Kormendy 1987b; Binggeli & Cameron 1991) and for background late-type galaxies. The intrinsic brightness distributions are then completely characterized by $m^T$, the scale length $h$, and the ellipticity $\epsilon$. For each image, we simulate exponential galaxies with various scale lengths and ellipticities, convolve them with a gaussian point-spread function of width $b_{\rm FWHM}$ and add the observed noise $\sigma_{\rm rms}$. For all exposures, the CCD read noise was small, so $\sigma_{\rm rms}$ is Poisson sky noise. We then run the same detection algorithm (see the next section) at the same surface-brightness threshold that we use for our data and we measure for each simulated galaxy the isophotal magnitude and the spatial moments of the light distribution $r_i = \int r^i I(\mathbf{r}) {\rm d}^2 \mathbf{r} / \int I(\mathbf{r}) {\rm d}^2 \mathbf{r}$ and $r_{jk} = \int x^j y^k I(\mathbf{r}) {\rm d}^2 \mathbf{r} / \int I(\mathbf{r}) {\rm d}^2 \mathbf{r}$, where $i, j, k = 0, 1, 2, \ldots$. Here $x$ and $y$ (in arcseconds) are the components of $\mathbf{r}$ in the west and north directions, and $r \equiv |\mathbf{r}| = (x^2 + y^2)^{\frac{1}{2}}$. By making the above analysis for a grid of $(m^T, h, \epsilon)$ values, we can compute the functions $m_I(m^T, h, \epsilon)$, $r_i(m^T, h, \epsilon)$, and $r_{jk}(m^T, h, \epsilon)$, where $i, j, k = 1, 2, 3, \ldots$. Only values of $\epsilon > 0.2$ are relevant in practice. Comparison of the measured parameters for intrinsically identical galaxies provides an estimate of the uncertainties in the transformation; these are due to noise fluctuations. The above functions are then combined and inverted to derive $m^T(m_I, r_i, r_{jk})$, $i, j, k = 1, 2, 3, \ldots$ and its uncertainty. The unknown intrinsic light distribution of the galaxy is now implicity encoded in the transformation through the moments of the light distribution. In practice, our measurements of simulated images suggest that the light distribution information is contained mainly in the first spatial moment $r_1$. For most galaxies, the dependence of $m^T(m_I, r_i, r_{jk})$ on the higher order moments is smaller than the uncertainties due to noise and non-exponential profiles (see Cowie et al. 1995 and Section 4.2.2). Therefore we simplify by writing the transformation as $m^T(m_I, r_1)$.

This function is computed for each image. An example is shown in Figure 2(a). Here $\Delta m = m_I - m^T(m_I, r_1)$ is the difference between the measured isophotal magnitude and the input true total magnitude. The uncertainty $\sigma(\Delta m)$ in $\Delta m$ is also shown. Both functions are complicated, but some general features are worth noting. For very compact galaxies with intermediate magnitudes $m_I \simeq 22$, $\Delta m$ becomes negative. That is, the isophotal magnitude of the galaxy is **brighter** than its true total magnitude. The reason is that the detection isophote contains local noise peaks. This effect is unimportant for bright galaxies, because noise peaks near the detection isophote do not contribute significantly to the isophotal magnitude. For very faint galaxies, it is overshadowed by the fact that much of the galaxy is below the surface brightness threshold. The fainter the brightness threshold, the more noise peaks are included in the isophote. The general trend is that the isophotal corrections get larger as the surface brightness of a galaxy gets lower, i.e., as its size gets larger at a given magnitude. The uncertainties are also larger for lower surface brightness galaxies, as shown in the lower panel of Figure 2(a). The sky brightness and seeing are important, too. If the sky brightness is high, its Poisson noise is high, and it becomes difficult to detect faint objects. As seeing gets worse, the galaxy light is spread out over a larger area and so more of the light falls below the detection threshold. When the seeing is extremely poor, the simulations suggest that $\frac{\partial r_1(m^T, h, e)}{\partial h} < 0$; i.e., intrinsically larger galaxies of a given magnitude have smaller measured isophotal sizes! If the seeing is worse than $1''$, for a typical night sky brightness, we find that this happens for $m^T > 22.8$, given values of $h$ that are typical for dwarf galaxies at $z \sim 0.02$. For this reason, we discard data where these effects are noticeable for galaxies other than extreme LSB galaxies (see Section 4.1).

Some regions of Figure 2(a) are not occupied by real galaxies that we detect. For example, bright objects with $r_1$ smaller than the seeing are not found; if they existed, such objects would occupy the front-left corner of Figure 2(a). Also, very faint, extended objects are not detected if their surface brightness is never brighter than the detection threshold; these would occupy the distant right of Figure 2(a). The completeness corrections describing the latter effect are presented in Figure 2(b). Two things are suggested by this figure. First, the completeness corrections depend strongly on surface brightness. Second, at a given surface brightness, the completeness drops from one to zero over a very short magnitude range (typically $\sim 1$ mag). What is measured in this transition region is the probability that a faint object is superimposed on a noise peak that is high enough so that the combined flux is higher than the detection threshold. The steep transition suggests that correcting for incompleteness is dangerous because the uncertainties in the correction would be huge. In this work, we present luminosity functions down only to magnitudes where galaxies that have scale lengths typical of cluster dwarfs are still detected at 100 per cent completeness. However, in addition to the above effect, which we shall call statistical incompleteness, there is further incompleteness due to crowding. That is, faint galaxies can remain undetected because they fall within the detection isophote of a much brighter galaxy. Corrections due to crowding can be important even for bright objects and will be addressed in Section 4.3.

### 4.2.2 Evaluation of the Technique

In this section, we estimate the systematic errors in our photometry for galaxies whose light profiles are not exponential. We also present arguments for and against the use of aperture magnitudes in this kind of study, and we outline our reasons for choosing isophotal magnitudes for measuring luminosity functions and aperture magnitudes for measuring colors.

Our method of calculating total magnitudes (Section 4.2.1) depends on the assumption that galaxies have exponential light profiles. We noted that this is a good approximation for late-type and dSph galaxies, but it is a poor approximation for ellipticals and for the bulges of disk galaxies. These tend to have de Vaucouleurs (1953) $\rho \propto r^{\frac{1}{4}}$ light profiles. We have applied our method to simulated de Vaucouleurs galaxies of known total magnitude and determined the systematic errors in measured $m^T$ values in the same way as we did for exponential galaxies in the previous section. The results are shown in Figure 3. This shows that for ellipticals at the distances of the clusters studied here, we underestimate the true luminosities by about 0.1 mag. This uncertainty is substantial; it is larger than the quadrature sum of all the other photometric errors described in Section 4.2.1. However, this is not a problem for the present study, for the following reasons.

(i) Even if **all** detected galaxies have de Vaucouleurs profiles, the errors introduced by assuming exponential profiles are still much smaller than the errors due to field-to-field variations in the background counts. This is least true at the faintest galaxies ($R \sim 25$) where the background variance is smallest, but at these magnitudes, the random errors due to noise are larger than the systematic errors due to non-exponential profiles (see Figure 3).

(ii) Most background galaxies at $20 < R < 25$ are late-type galaxies. Very few are ellipticals. Local late-type galaxies have approximately exponential profiles (e.g., Freeman 1970). A number of late-type galaxies at high $z$ show morphological peculiarities (Cowie et al. 1995), but any resulting systematic errors in our photometry are likely to be similar for our background and cluster fields. Therefore these errors will cancel out during background subtraction. We only need to worry if the **cluster** galaxies detected at $20 < R < 25$ have light profiles that deviate significantly from exponentials. But the results of Faber & Lin (1983), Kormendy (1987b), and particularly Binggeli & Cameron (1991), show that the vast majority of dSph and dIrr galaxies locally and in Virgo have exponential profiles. No true ellipticals exist this faint (they would be at least one magnitude fainter than M32) locally or in Virgo (Binggeli 1987; Binggeli et al. 1988). Globular clusters also are low-luminosity systems that do not have exponential profiles, but at the distance of these clusters, their scale lengths are too small to be resolved. Therefore it seems safe to assume that the dwarf galaxies that we find have exponential brightness profiles.

(iii) In (ii), above, we noted that most galaxies with nonexponential light profiles will be distant ellipticals or peculiar star-forming galaxies at high $z$. In both cases, they will be much farther away than the cluster we are studying, so their apparent sizes will be small. Figure 2 then implies that we underestimate the luminosity by less than Figure 3 suggests.

Is it better to use aperture magnitudes or isophotal magnitudes to estimate the total magnitudes? Figure 4 shows how the two methods compare for simulated galaxies. It suggests that the aperture magnitudes are better for faint, compact galaxies, but they are extremely poor for dwarf galaxies at $z = 0.02$ and $R \sim 20$ because the scale lengths of such galaxies are too big. Because we need accurate photometry on these galaxies to measure the luminosity function, we do not use aperture magnitudes. The errors are slightly bigger for the isophotal method at $R > 22$, but they are small compared to those introduced by the field-to-field background count variations. We therefore would gain little if we had slightly smaller photometric errors at the faintest magnitudes.

On the other hand, we use aperture magnitudes for measuring galaxy colors. Then we are sure that we measure light from the same region of the galaxy in both bandpasses. Missing substantial fractions of the light for diffuse galaxies is not a problem unless the galaxies have huge color gradients.

In summary, we use corrected isophotal magnitudes as described in Section 4.2.1 for computing the luminosity functions. Aperture magnitudes are calculated for all the galaxies, but these are used only to measure colors. We accept the random errors in the lower panel of Figure 2(a). We note that there may be systematic errors in our photometry if dwarf galaxies in the clusters have nonexponential profiles. However, for de Vaucouleurs galaxies, these systematic errors are negligible compared to the statistical uncertainties resulting from field-to-field variations in the background counts. Therefore our measurements of luminosity function are insensitive to systematic errors in the photometry. Furthermore, local galaxies whose apparent magnitudes would be $20 < R < 25$ at $z = 0.02$ almost invariably have exponential profiles, so we do not regard the existence of a nonexponential population as likely in the first place.

**4.3 Measurements of the galaxy number counts**

We use the Faint Object Classification and Analysis System (FOCAS) of Jarvis & Tyson (1981; see also Valdes 1982, 1989) to detect objects and to compute their basic photometric quantities. An object is detected if 15 or more contiguous pixels exceed the local sky by more than $3\sigma_{\rm rms}$; this is equivalent to using a detection isophote of $\sim 27$ mag arcsec$^{-2}$ for both our $B$- and $R$-band images. Multiple objects within a single detection isophote are identified by searching for multiple brightness maxima; these are split into individual objects. Objects are then classified based on their morphology relative to several reference PSF stars in the field. Valdes (1982, 1989) presents the classification terminology; the main classes are "stars", "galaxies" (these have broader profiles than the PSF), "diffuse" (these have even broader profiles than galaxies), "long" (these have highly asymmetric profiles), and "noise" (these are smaller than the PSF). The values of $m_I$ and $r_1$ are then computed following a local sky subtraction, and a catalog is output containing the classification, the values of $m_I$ and $r_1$, the isophotal area, and the coordinates of each object. We typically detect 1000–2000 objects per image. A similar catalog was derived with the splitting algorithm turned off; comparison of the catalogs provides a list of the objects that were part of a bigger object in the initial detection pass. The following modifications are then made to the catalog:

(i) From the simulations in Section 4.2, we decide on the faintest magnitude $m_L$ to be included in the number counts. This is chosen as the faintest magnitude at which dSphs at the distance of the cluster are detected with 100 per cent statistical completeness and with $\sigma(\Delta m) < 0.5$. Dwarf spheroidals are assumed to have $M_B \approx M_R + 1.5 \approx -9.8 \log_{10} h - 16.6$, based on scale lengths given by the ridge line of the Local Group fundamental plane correlations (Kormendy 1987b) and on colors from Caldwell (1983). We also require that $\frac{\partial r_1(m^T,h,e)}{\partial h} < 0$ for all $m^T < m^T(m_L, r_1)$, where $h$ is less than or equal to its value for a typical dSph galaxy at the redshift of the cluster; otherwise the simulations show that deviations from exponential profiles can result in unreliable values of $m^T$. Objects whose measured $m_I$ and $r_1$ suggest a value of $m^T$ that is fainter than $m_L$ are deleted from the catalog. This leaves typically 60–90 per cent of the 1000–2000 objects detected at the $3\sigma$ isophote.

(ii) All remaining objects that were not classified as "stars" or "galaxies" were examined individually and a decision was made as to whether or not they should be included. Most were "diffuse"; these are probably faint, extended galaxies, so they were retained. A few noise peaks (classified "noise") and the diffraction spikes of bright stars (classified "long") had to be rejected in some images. At this stage, we also rejected objects within $\sim 1''$ of the field edges, including all objects whose centers were not within the image.

(iii) All objects that were part of a larger object in the original detection pass and that were subsequently split into more than three objects were investigated individually by eye. This allowed us to remove several noise peaks that looked like faint stars or galaxies in the halos of bright stars and galaxies. Here the effective sky brightness and noise are higher than average, so more noise peaks get detected. Also, several galaxies that had numerous bright spots were reassembled from their split components at this stage. This was particularly a problem for extreme LSB galaxies, where the surface brightness was so low that noise peaks caused the splitting algorithm to recognize the galaxy as an assemblage of several smaller objects. Even though this process is extremely time-consuming, we feel that it is worthwhile, because it gives us more confidence that the objects in our catalog are truly individual galaxies or stars. For isolated objects, automated detection programs like FOCAS tend to perform well, but for merged objects, they have difficulty in making decisions about the true nature of split objects. This is mainly because there is no straightforward splitting algorithm. Experiments with simulated images where the split nature of the objects is known *a priori* suggest that the eye does better. Since 20–40 per cent of the objects we detect typically fall into this category, we feel justified in investing a significant effort to make these judgments by eye. Even then, there are $\sim 10$ objects per field for which it is impossible to tell if the object is a single galaxy or two galaxies. However, these cases are rare enough that resulting errors are much smaller than the counting statistics. Finally, FOCAS is used to compute the photometric parameters of all objects that we have decided are individual stars or galaxies.

One might expect that $\sigma(\Delta m)$ is larger for split objects than for isolated objects, but our simulations suggest that this effect is not severe in most cases. However, near very bright stars or galaxies, the errors do grow. Problems are caused both by the rapidly varying background and by its associated noise, as well as by fluctuations in the background due (e.g.) to shells or spiral structure inside bright galaxies. Our solution is to exclude such regions from the number counts. Typically, we discarded regions where the additional sky brightness due to the halo of the bright object was comparable to the night sky brightness: $\sigma(\Delta m) > 0.5$ in such regions is not rare for objects that would have $\sigma(\Delta m) < 0.1$ in the rest of the image. By discarding areas covered by bright galaxy halos, we give up sensitivity to variations in the number counts on scales much less than 20 kpc from the centers of objects like NGC 708 and NGC 7626. However, there is evidence that the number counts in such regions are dominated by globular clusters, anyway (e.g., see B95).

At this stage, we have a catalog of objects. They are labelled by FOCAS as "galaxies" or "stars", but at intermediate and faint magnitudes, these classifications can be unreliable. For example, many galaxies may be compact or nucleated and so can have scale lengths that are smaller than the seeing. These look like stars. To see whether confusion with real stars can be important, we need to estimate the contribution of faint stars to our counts. To estimate this, we assume that the shape of the Galactic stellar luminosity function (SLF) is invariant at $20 < m < 25$. We adopt the SLF shape by Jones et al. (1991) between $20 < m < 25$ and normalize it by the numbers of stars we detect brighter than $R = 20$ or $B = 21$ (there were 20–70 such stars in each image). For such bright stars, the FOCAS classifications are 100 per cent reliable (this was checked using a simple PSF-fitting algorithm; saturated stars were identified by eye). We then compute the number of faint stars as a function of magnitude. The total stellar contamination proves to be less than 10 per cent at faint magnitudes in most of our images. Therefore the uncertainties from possible misclassification of galaxies as stars are much smaller than the errors from galaxy counting statistics. They can safely be neglected.

From $m_I$ and $r_1$, we then compute $m^T$ and its uncertainty $\sigma(m^T) = \sigma(\Delta m)$ using the methods of Section 4.2. We correct for stellar contamination using the measured stellar contribution brighter than $R = 20$ or $B = 21$ and the predicted contribution fainter than this. We bin the cluster counts in one-magnitude bins and the background counts in half-magnitude bins. Finally, we compute the number-count-magnitude relation in units of number of galaxies per magnitude per square degree by dividing the number of galaxies in each bin by the surveyed area. The surveyed area is smaller than the total field size given in Table 2 because we omitted, in decreasing order of importance, (a) regions surrounding large galaxies where the measured magnitudes are unreliable because of varying noise and features in the galaxy halos (see above), and (b) regions that are too close to the edges of the field. Also, (c) we corrected for crowding. Simulations suggest that at faint magnitudes, the total isophotal area contained within bright objects whose isophotal area is greater than 1000 contiguous pixels is a reasonable estimate of the the area that is lost due to crowding at faint magnitudes (typically 2% of the total field area). At bright magnitudes, the correction for crowding is much smaller than the errors from galaxy counting statistics; adopting the same number here too is therefore adequate. Finally, the luminosity function was corrected for Galactic extinction using the maps of Burstein & Heiles (1982) and the color conversions of Cardelli et al. (1989).

These techniques were tested on simulated fields. An example is shown in Figure 5, where half-magnitude bins are used. The last point represents $m_L + 0.5$; again we stress that for our clusters we will not present data fainter than $m_L$ (for the background fields presented in Section 4.4, the $m_L + 0.5$ points are shown, however). Contributions to the errors come from counting statistics and from $\sigma(m^T)$; for the $m_L + 0.5$ point, the uncertainty due to the statistical completeness correction is also included. For most of our simulated images, counting statistics dominate the uncertainties except at $m > m_L - 2$, where $\sigma(m^T)$ dominates. Measurements of the faint-end slope of the luminosity function depend strongly on this magnitude range, so taking account of uncertainties from $\sigma(m^T)$ is important.

### 4.4 The background fields

The analysis described in the previous two sections was performed on all of the background fields listed in Table 2. Figure 6 shows the results, together with those from other authors. In the range $21 < m < 25$, our $R$-band number counts are well fit by the equation

$$\log_{10} N = 0.376R - 4.696 , \qquad (2)$$

and the $B$-band counts are well fit by

$$\log_{10} N = 0.488B - 7.702 . \qquad (3)$$

Here $N$ is the number of galaxies per square degree per half-magnitude interval; $B$ and $R$ represent the $B$- and $R$-band total magnitudes, equivalent to $m^T$ in the previous section. We use the above equations to estimate the background contribution to the counts in our cluster fields. Brighter than $m = 21$, the systematic deviations from the measured counts in Figure 6 that we would expect if use these equations are far smaller than the uncertainties from counting statistics; therefore we can safely use these equations over the entire magnitude range (for most our clusters, the luminosity function is poorly determined at $m \ll 21$ anyway).

Equations (2) and (3) give the **mean** number counts for background fields selected at random. Also important is the **variance** in the counts; this is the source of the greatest uncertainty in the cluster luminosity functions that we derive. We use the results of B95 to determine this variance, as follows. From their Table 2, their Figure 4, and their equations (1–5), we compute the value of $\left(\frac{\Delta N}{N}\right)_{\text{bkg}}$, the fractional standard deviation in the background number counts for their fields, expressed as a function of $R$ magnitude. We then fit a second-order polynomial to this function; we correct for the slight difference between their field size and ours assuming Poisson statistics, and we adopt the result as our estimate of the field-to-field variation in $R$-band background counts.

We use the results of B95 because they observed more background fields than we did. They have four fields. We have three, but one of ours is at low Galactic latitude, so the uncertainty in the counts is slightly larger due to uncertainty in the Galactic extinction. Their field size is also larger than ours. We do not combine our number counts with theirs because of uncertainty in the relative zero point. But we do note that the variance in our counts is consistent with that observed by B95. This is reassuring.

Since we do not have a direct measurement of the variance in the $B$-band counts, we derive this from the $R$-band counts. We therefore note that background galaxies in the range $21 < B < 25$ have a distribution of $B - R$ colors that is strongly peaked at 1.5 (Driver et al. 1994a), and we scale the variance in the $B$-band counts accordingly.

Figure 7 shows the values of $\left(\frac{\Delta N}{N}\right)_{\text{bkg}}$ that we compute. We assume that these functions and equations (2) and (3) give a complete characterization of the distribution of background counts for our cluster fields. Note that the uncertainties $\left(\frac{\Delta N}{N}\right)_{\text{bkg}}$ do not necessarily have a Gaussian distribution; large-scale structure at $z \geq 0.1$ may skew the probability spectrum toward high $N$. This will be discussed further in Sections 5 and 6.

## 4.5 Measurement of the luminosity function

The analysis described in Sections 4.2 and 4.3 was performed on the cluster fields listed in Table 2. The resulting number-count–magnitude plots are given in Figure 8. The error bars represent the combined uncertainties from counting statistics and from uncertainties in the isophotal magnitude corrections; counting statistics dominate for all but the faintest two or three points in each plot. Also shown in Figure 8 are the mean background counts for each field given by equations (2) and (3) and corrected to the larger bin size.

Figure 8 shows that for many of our cluster fields, the number counts are systematically above the background. Our interpretation is that the excess objects are associated with the cluster. (Whether they are dwarf galaxies or globular clusters is considered in the next section.) The luminosity function of the excess objects is then computed for each frame by subtracting the background.

Additional sources of uncertainty in the background counts that may affect the results in a systematic way are the effects of gravitational lensing by the cluster dark matter and the obscuration of background galaxies by dust in the cluster. We follow B95 in neglecting these effects for clusters at $z \sim 0.02$. The reasons are:

(i) Gravitational lensing of the background population is negligible because (a) the cluster surface density $\Sigma \ll \Sigma_c$ at all radii, and (b) from equations (2) and (3), $2.5 \frac{d \log_{10} N}{dm} \approx 1$ for the background counts. These two effects then combine to result in a multiplicative change in the background counts due to gravitational lensing

$$f_{\rm lens} = |1 - \frac{\Sigma}{\Sigma_c}|^{1 - 2.5 \frac{d \log_{10} N}{dm}} \qquad (3)$$

that is very close to one at all magnitudes. This equation is valid here because most background galaxies are very much farther away than the clusters; for these galaxies, $\Sigma_c(z) \approx \Sigma_c(\infty)$. These numbers suggest that the uncertainty introduced in assuming $f_{\rm lens} = 1$ is much smaller than $\left(\frac{\Delta N}{N}\right)$ and can be safely neglected.

(ii) The observations of elliptical galaxy colors by Ferguson (1993) suggest that the total extinction $E_{B-V}$ introduced by the clusters in his sample (which includes A262 and Pegasus) is less than 0.05 magnitudes. This translates into an uncertainty far less than $\left(\frac{\Delta N}{N}\right)_{\rm bkg}$.

As B95 point out, neglecting both of the above effects for clusters at $z \sim 0.02$ causes us to overestimate the background and therefore to underestimate the cluster counts. Globular cluster contamination and large-scale structure at $z \geq 0.1$ are additional sources of uncertainty in the background; these are more worrying because ignoring them can cause us to underestimate the background counts. These problems will be addressed for each cluster in the following section.

## 5 RESULTS

The results are presented in Figures 9 through 12. Here $M$ is the absolute magnitude, equivalent to the total apparent magnitude $m$ (or $m^T$ in Sections 4.2 and 4.3) minus the distance modulus. The error bars now represent the quadrature sum of the uncertainties in the subtracted background counts (which dominate), uncertainties in the aperture corrections, and counting statistics.

### 5.1 Abell 194

The luminosity function of A194 is shown in Figure 9. Also shown is the radius-magnitude correlation for the stars and galaxies that we detect. It is clear that the uncertainties in $\phi$ are too large for a convincing measurement of $\alpha$. Table 3 gives the value of $\alpha$ obtained from a power-law fit to the data in Figure 9(d) including their $1\sigma$ uncertainties. As outlined in Section 4.5, we assumed a variance in the background (Figure 7) corresponding to that measured by B95. In Table 3, we also show how much weaker the constraint on $\alpha$ would be if the true variance were larger, as it might be if large-scale structure at $z \geq 0.1$ skewed the high-$\left(\frac{\Delta N}{N}\right)_{\text{bkg}}$ end of the background counts distribution. The results are presented in terms of $\eta$, where (in obvious notation)

$$\eta = \left(\frac{\Delta N}{N}\right)_{\text{bkg,true}} \bigg/ \left(\frac{\Delta N}{N}\right)_{\text{bkg,B95}} . \qquad (4)$$

If $\eta \geq 2$, then $\alpha$ is not constrained even at the $1\sigma$ confidence level. However, if $\eta = 1$, then $\alpha < -1.6$ at the $1\sigma$ confidence level for $-15.88 < M_R < -9.88$. This is substantially steeper than $\alpha \simeq -1.35$ found for Virgo by SBT85 (although our measurements probe a different magnitude range than theirs). This result depends critically on the assumptions that $\eta = 1$ and that globular cluster contamination in the faintest two data points is small (brighter than $M_R = -12$, globulars are rare). We do not regard globular cluster contamination as a serious problem for A194 (i) because most objects appear to be resolved, (ii) because the brightest galaxies in the cluster are, at $M_R \sim -21$, substantially fainter than the sort of brightest cluster galaxy (e.g., M87, McLaughlin et al. 1994; NGC 4874, B95) whose globular cluster population dominates the number counts in its halo, and especially (iii) because the galaxies in the faintest two bins in Figure 9(d) are not clustered around the giant galaxies. Globular cluster contamination will be addressed more quantitatively for the other clusters, where we convincingly detect excess objects and hence can measure $\alpha$.

Note from the dot-dashed line in Figure 9(c) that most of the faint objects are smaller than local dSphs would be if moved to A194. If these are cluster galaxies, they are more compact than local dSphs. They may be nucleated dSphs, or they may be a different kind of object.

Figure 9(c) also suggests that extremely low surface brightness galaxies (those above the solid line in the figure) do not contribute more to the galaxy luminosity function at fainter magnitudes. Although we find some objects that are considerably more diffuse than anything seen in the field, they are rare. Of course, there may be a substantial population of galaxies that are so far above the solid line that we do not find them at all. We must always be aware of this caveat and regard the observed number counts, here and in other clusters, as lower limits. The possibilty of such a population has long been recognized (Arp 1965, Disney 1976) and remains a potential worry for all studies of this type (see also Phillipps et al. 1988).

From Figure 9(a), the relative $B$ and $R$ luminosity functions are not determined well enough to constrain the colors of the dwarf population. However, we note that the mean color of the extreme LSB galaxies (i.e., those more diffuse than anything observed in the field) is $B - R = 1.3$, close to the mean color of the giant galaxies in the cluster (Chapman et al. 1988).

We cannot constrain $\phi$ in A194 is because the dwarf galaxy density is too low. This could be because the dwarf-to-giant ratio is low or because the total galaxy density is low. Our observations do not distinguish between these possibilities.

**5.2 Abell 262**

Figure 10 shows the luminosity function of A262. It is immediately apparent that these results do constrain the luminosity function and that it is steep. Table 4 lists $\alpha$ as derived from a power-law fit, $\alpha = -1.84^{+0.28}_{-0.36}$ (1$\sigma$ uncertainties) and the effect of varying $\eta$. Results are also given for Field I by itself; since it contains most of the excess galaxies, it provides slightly stronger constraints: $\alpha = -1.90^{+0.22}_{-0.28}$. As mentioned in Section 3, A262 is the richest cluster in our sample and the one that is most similar to Virgo. It is therefore intriguing that its luminosity function is so close to that suggested by IBM88 for Virgo. Furthermore, the turn-up in the luminosity function that we see in A262 occurs at a similar absolute magnitude as the turn-up that they hypothesize in their Figure 10.

However, we need to check that these results are not due to errors in the assumed background counts or to globular cluster contamination.

### 5.2.1 Globular Cluster Contamination

The central galaxy in A262, NGC 708, has $M_R \sim -24$ and therefore may well have a large globular cluster population. However, most of the faint objects that we detect appear to be resolved [Figure 10(c)]. Moreover, they are uniformly distributed, especially in Field I. This would suggest that the globular cluster contamination is small. Confirming this, we note that the luminosity function is unchanged if we exclude all objects within 50 kpc of NGC 708 (the only significant difference is that the lower logarithmic error bar on the faintest point gets bigger by a factor of two; much of this difference is due to poorer counting statistics). The globular cluster contribution to the total counts is likely to be overwhelmingly dominated by clusters within 50 kpc of NGC 708 (see McLaughlin et al. 1994 for M87, which is more luminous than NGC 708 and which still has most of its globulars within 50 kpc of the galaxy center). Therefore we conclude that the globular cluster contribution to $\phi$ is small.

Table 4 also shows how the measurement of $\alpha$ would change if we removed the faintest point in Figure 10 (a) and (d), i.e., the one that is most affected by globular cluster contamination. As long as $\eta \simeq 1$, the results still imply that $\alpha \approx -1.8$.

### 5.2.2 Contamination from an anomalously high number of background galaxies

A strategy like that used in Section 5.2.1 was used to investigate whether or not a background cluster at high redshift could contribute the excess objects that we found. To check this, we found the center of the galaxy distribution in Field I, i.e. $\mathbf{r}$ such that $\Sigma|\mathbf{r}-\mathbf{r}_i|$ for all objects is minimized, counting each object as a single data point. This calculation is very insensitive to the stellar contamination. We then recomputed the luminosity function, excluding the half of Field I that is centered on $\mathbf{r}$. No significant difference in $\phi$ was found. The excluded region had a radius of $2'.5$; this is substantially larger than the core radius of a cluster at or beyond the predicted median redshift of the faintest background galaxies, $z = 0.4$ (Cowie et al. 1991). A cluster at $z \simeq 0.1$ would have a core radius approximately equal to our field size and might affect our counts, but such a cluster should have been found in the Sakai et al. (1994) study of the velocity distribution of galaxies in A262. Also, A262 is on a filamentary ridge of the Pisces-Perseus supercluster, so we expect that foreground and background contamination from supercluster members is small.

The most straightforward interpretation is therefore that we have detected a population of dwarf galaxies with $\alpha \approx -1.8$. At $R \simeq 24$, these are similar in size to dSphs that satisfy the Local Group$-$Virgo fundamental plane correlations [Figure 10(c)]. At $R \sim 21$, the A262 galaxies are slightly more compact. However, the scatter in Figure 10(c) is larger than this difference. Figure 13 shows the color distribution of the excess galaxies; it suggests that they are redder than the background galaxies (mean $B - R$ color $\simeq 1.7$). This also is consistent with dSph galaxies – if the faint galaxies were dwarf irregulars, the peak in the histogram in the lower panel of Figure 13 would be shifted to much bluer colors ($B - R \simeq 1.0$).

As in Abell 194, we find a number of LSB galaxies [those above the solid line in Figure 10(c)], but they do not dominate the luminosity function at faint magnitudes.

It is also apparent from Figures 10 and 12 that the galaxies we detect are mostly confined to Field I. Therefore the galaxy density drops significantly on a scale that is close to our field size ($\simeq 180$ kpc). This is the core size of a typical small cluster, so the difference between Fields I and II is not surprising.

In summary, we find an excess population of galaxies above the background in A262, but only in a 200 kpc region around the cluster center. The colors and scale lengths of the excess galaxies are similar to those of dSph galaxies. This suggests that dSph galaxies overwhelmingly dominate the faint end of the luminosity function, as they do in Virgo (Binggeli 1987). This is not surprising, given the similarities between the two clusters (Section 3). Fainter than $M_R = -14$, the dSph galaxies have a steep luminosity function, with $\alpha \simeq -1.8$.

### 5.3 Pegasus

The luminosity function for the Pegasus Cluster is presented in Figure 11 and Table 5. The format is the same as for the previous clusters. In Field I, we observe a large excess of objects fainter than $M_R = -11$ and relatively few that are brighter than this. Moreover, $\phi$ is steep, but $\alpha \simeq -2.5$ for $\eta = 1$ is determined primarily by the two points at $M_R \simeq -10$. Also, Field I is between the giant ellipticals NGC 7619 and NGC 7626 near the center of the cluster. Finally, the $M_R \gtrsim -11$ objects are unresolved. We therefore interpret the faint objects as globular clusters. The luminosity function of galaxies in Pegasus is essentially unconstrained. As in A194, the galaxy density is too low.

### 5.4 NGC 507 Group

The luminosity function for the NGC 507 Group is presented in Figure 12. Figure 12(f) shows that the luminosity function is better constrained than for any of our other clusters. This is partly because we surveyed four fields and partly because the dwarf galaxy density is high. Contamination by background clusters of galaxies and by globular clusters belonging to NGC 507 are not serious problems for the same reasons as in A262. It is not surprising that globulars are not a problem: NGC 507 is a late-type galaxy, and these have systematically lower specific globular cluster frequencies than elliptical galaxies (Zepf & Ashman 1993). Furthermore, the strongest excess of objects is in Field IV, well away from NGC 507. But an excess is seen in the other fields, too. This excess is strong at intermediate magnitudes ($M_R \simeq -14$) as well as at faint magnitudes (see Figure 1: Field IV is visibly richer at intermediate magnitudes than any of the other fields). The giant galaxies are more spread out in this group than in A262 (Sakai et al. 1994); our results suggest that the same is true for the dwarf galaxies. Like A262, this cluster is on a filamentary ridge of the Pisces-Perseus supercluster. Any background clusters at $z \simeq 0.1$ would probably have been identified by Sakai et al. (1994).

Table 6 lists the derived values of $\alpha$. Note that a power-law fit fails at the $1\sigma$ confidence level if $\eta \leq 2$. This suggests that the curvature of the luminosity function is a real effect. This is not worrying; it is well-known that there is no *a priori* reason why the luminosity function should be a power-law (or any other analytic function). However, it is clear that the luminosity function is steep, with $\frac{d \log \phi(L)}{d \log L} \approx -1.6$. Fainter than $M_R = -12$, the data suggest that $\frac{d \log \phi(L)}{d \log L} \approx -2$.

The galaxies we detect have scale lengths that are slightly smaller than for typical Local Group and Virgo dSph galaxies moved to the distance of the cluster. The distribution of the colors of the excess galaxies found in Field I is presented in Figure 15. The counting ($\sqrt{N}$) statistics are poorer here than in Figure 14, so we are unable to make rigorous interpretations as to the types of dwarf galaxies we are finding. However, we note that the color distribution is somewhat redder than we would expect if most of the dwarfs were irregulars.

## 6 DISCUSSION

Our results are therefore as follows. For A262 and for the NGC 507 Group, there is definitive evidence for a steep luminosity function. For A194, the results suggest a steep luminosity function but are significant only at the $1\sigma$ level. For these three clusters, $\phi \propto L^\alpha$ has a faint-end slope of $\alpha \simeq -1.8 \pm 0.3$. Pegasus has too low a galaxy density for us to measure $\alpha$.

It has been known for more than a decade that Virgo has a steep luminosity function, with $\alpha$ at least as small as $-1.35$ (SBT85) and perhaps as small as $-1.7$ (IBM88). That many clusters might have steep luminosity functions at low $L$ has only recently been recognized because of a number of claims similar to ours (Driver et al. 1994b, De Propris et al. 1995)[1]. Now we add three more clusters to the list. It is intriguing that there are also theoretical hints that unevolved environments may have steep luminosity functions (e.g., Babul & Ferguson 1996). As with any observational result that is new and somewhat surprising, it is necessary to be particularly careful about caveats. How much confidence should we have in our results?

Uncertainties in $\alpha$ come mainly from uncertainties in the background counts. Therefore the critical question is: How sure are we that $\eta \approx 1$? In the $R$ band, B95 surveyed four background fields and we surveyed three. We did not combine the two data sets because of possible zeropoint differences, but the statistics implied for the parent population of background fields is not consistent with the hypothesis that large fluctuations explain the excess counts in *all* of our clusters. This conclusion is based on the assumption that the statistics are Gaussian. Large-scale structure at $z \geq 0.1$ may skew the high-$\left(\frac{\Delta N}{N}\right)_{\mathrm{bkg}}$ end of the count distribution in a way that is difficult to quantify. However, we believe that this is not likely to explain the excess counts because all three clusters have been studied spectroscopically (Chapman et al. 1988; Sakai et al. 1994), and these studies should have found background clusters at $z \sim 0.1$. More distant clusters are not a problem because the galaxies we detect are too smoothly distributed in our fields.

Additional support for our results comes from the fact that our photometry technique takes the surface brightness of the galaxies into account. All galaxies in Figures 8 – 12 are bright enough that the luminosity functions do not depend on completeness corrections except for those due to crowding.

---

[1] These papers found steep luminosity functions at magnitudes somewhat brighter than those studied here. They stimulated great interest. However, the results may suffer from serious technical problems and not be as secure as originally thought (Trentham & Kormendy 1996).

– 30 –

We therefore suggest that steep luminosity functions at $M_R \sim -12$ may be a universal feature of young clusters. The relevance of this result to the field is uncertain, but if the galaxies in our clusters have not experienced significant galaxy-galaxy interactions, then they may be representative of the field as well. If the total luminosity function in the field is steep, then Tully's (1988) conclusion that gas-rich field galaxies have $\alpha \simeq -1$ suggests that the faint galaxies which are so numerous in the field are not gas-rich (however, see Marzke et al. 1994). Presumably they are dSph galaxies. Many authors (e.g., Kormendy 1988, 1990; Babul & Rees 1992; Babul & Ferguson 1996) have proposed that faint or even invisible gas-poor dwarfs may be very common.

We can also combine our results and similar ones for giant galaxies (Kirshner et al. 1983) with the dark matter scaling laws (Kormendy 1988, 1990) to estimate the cosmological density of the dark matter that is contained in dwarf galaxy halos. If we assume that the dwarf-to-giant ratio is the same in the field as it is in Virgo and if $\alpha = -1.8$, then $\Omega_{\mathrm{dwarfs}} \approx 0.01$ for a galaxy mass range of $10^6$ to $10^{11}$ $M_\odot$. It is $\Omega_{\mathrm{dwarfs}} \approx 0.02$, for a mass range of $10^4$ to $10^{11}$ $M_\odot$. In this calculation, we assumed that the dSph luminosity function has a Schechter-like (exponential) turnover at the bright end (Ferguson & Sandage 1991 justify this; we use their Virgo value of $M^*_{\mathrm{dSph}} = -16.5$ for the turnover magnitude and assume that an $M^*_{\mathrm{dSph}}$ dSph galaxy has a global mass-to-light ratio, including dark matter, of 15). The above values of $\Omega_{\mathrm{dwarfs}}$ are substantially smaller than the amount of dark matter implied by dynamical measurements on scales of $\sim 10$ Mpc (Shaya et al. 1995) or the amount of cluster dark matter ($\Omega \approx 0.2$ from the Coma inventory argument of White et al. 1993). This suggests that dwarf galaxy halos do not contribute significantly to the cosmological mass density. There are two caveats: (1) Tidally disrupted remnants of even smaller galaxies may be important, even if they are smoothly distributed today. (2) If the fraction of mostly dark galaxies that get discovered is $\ll 100$ per cent for the smallest dwarfs, then their halos may be more important than we think. The above result also depends on the assumption that the dark matter scaling laws observed locally are universal. It is nonetheless reassuring that we do not find more halo dark matter than the dynamics imply. This would have been the case if $\alpha$ were significantly more negative than we found.

The luminosity function of dwarf galaxies in young environments provides an important test of models of galaxy formation and evolution. In particular, our measurements probe the mechanisms that control star-formation efficiencies in low-mass galaxies. For example, the model presented in Figure 14 of Babul & Ferguson (1996) predicts a luminosity function that is steeper ($\alpha \simeq -2.7$) than can be reconciled with our observations. This discrepancy may be explained if many of their dwarfs fall below our surface brightness detection threshold or if dwarfs are destroyed in the early stages of cluster formation. Our results also appear inconsistent with the no-merging model of White & Kauffmann (1994), but for the opposite reason – our luminosity function is too steep at the faintest magnitudes. However, the luminosity functions suggested by some of their models (those with low dynamical friction merging timescales) are consistent with our data. These are just two examples among many: the comparison of dwarf galaxy luminosity functions with semi-analytic models should be a vigorous area of study as the observations improve and the constraints on $\phi$ become stronger. Also, measurements deeper than ours may reveal a turnover in $\phi$ if formation of the smallest galaxies is suppressed by photoionization by the UV background (Efstathiou 1992). No turnover is observed brighter than $M_R = -10$ in A262 and in the NGC 507 Group.

In this work, we fitted power-law luminosity functions to our data. We succeeded for A262 but failed for the NGC 507 Group. We used this simple approach so that our results can be related more easily to previous work, but we stress that there is no physical motivation for any analytic fitting function (for a discussion of the motivations and techniques for fitting functional forms to $\phi$, see Ferguson & Sandage 1991). It is therefore not worrying that a power law fails to fit $\phi$ in the NGC 507 Group. Indeed, it is encouraging that our data are good enough that we can measure the curvature of the luminosity function. Since the derivation of $\alpha$ dependends on the assumption of a power law, our quoted values for the other clusters should be viewed with caution. A better approach would be to make direct comparisons (in terms of some goodness-of-fit parameter) between observed and model luminosity functions. As data and models become more accurate, such comparisons will probably become the normal way of presenting the results. More generally, Efstathiou has pointed out (see Phillipps 1994) that $\phi$ is probably better represented as a multivariate function of several structure parameters (like radius, luminosity, and a characteristic velocity). We cannot adopt this approach yet because we lack velocity data. However, with the advent of 8- and 10-meter telescopes, it will soon be possible to obtain enough kinematic measurements to make this approach productive.

Observations of dwarf galaxies in clusters are progressing rapidly. In particular, large-area surveys to $R \sim 26$ are now possible due to the advent of $4000 \times 4000$ and $8000 \times 8000$ pixel CCDs. This means that large enough areas can be surveyed so that the uncertainties in background counts can be reduced. Also, multi-object spectroscopy can provide large numbers of redshifts for faint galaxies. This will allow us to determine luminosity functions independently of background subtraction (e.g., Biviano et al. 1995), although not to the faint limits studied here.

## ACKNOWLEDGMENTS

I thank John Kormendy for taking the observations of Background Field 3 at the CFHT in January 1995, for many helpful discussions, and for detailed comments on the manuscript. Helpful discussions with Len Cowie, Brent Tully, Richard Wainscoat, Lev Kofman, and Xerxes Tata are gratefully acknowledged. This research has made use of the NASA/IPAC extragalactic database (NED) which is operated by the Jet Propulsion Laboratory, Caltech, under agreement with the National Aeronautics and Space Administration.

**FIGURE CAPTIONS**

**Figure 1.** $R$-band images of two background fields at different galactic latitudes $b$ and of all cluster fields surveyed. North is up and east is to the left in all images. Field sizes, seeing, and exposure times are given in Table 2. The faintest objects that are easily visible have $R \approx 22$. Giant galaxies in these frames include NGC 538 [(c), lower right], NGC 541 [the bright galaxy in the center left of frame (d)], Arp 133 (the peculiar galaxy northeast of NGC 541), NGC 535 [near the center of frame (d)], NGC 708 [the brightest and central elliptical galaxy of Abell 262; center right of (e)], NGC 705 [at the far right of frame (e), southwest of NGC 708], NGC 703 [extreme left center of frame (f)], NGC 704 [the bright pair of galaxies at the bottom left of frame (f)], NGC 7626 [the bright galaxy in frame (g); NGC 7619, the other bright central galaxy in Pegasus, is just outside this field to the west], NGC 507 and 508 [the brightest galaxies in (i)], NGC 503 [southeast corner of frame (l)], and 01205+3305 [the interacting group of galaxies towards the lower left of frame (l)].

**Figure 2.** (a) The upper panel gives an example of the function $\Delta m(m_I, r_1)$, the aperture correction applied to galaxies with $3\sigma$ isophotal magnitudes $m_I$ and first-moment light radii $r_1$ (arcsec). These were the corrections for the $R$-band image of Field 1 of Abell 194. The scale was $0''\!.22$ pixel$^{-1}$; the FWHM of the seeing was $0''\!.79$, the sky brightness was 21.0 $R$ mag arcsecond$^{-2}$, and the rms sky noise was 25.9 $R$ mag arcsecond$^{-2}$. The lower panel shows the uncertainty $\sigma(\Delta m)$ in this function. (b) Detection completeness for simulated galaxies with exponential light profiles in the the same image. The solid line is for galaxies with scale lengths characteristic of late-type, giant galaxies at $z = 0.02$. The dashed line is for galaxies with scale lengths characteristic of dSph galaxies at $z = 0.02$ (see text).

**Figure 3.** Errors made when the isophotal corrections of Section 4.2.1 are applied to galaxies that have $r^{\frac{1}{4}}$ brightness profiles. (a) The solid line shows the difference between the true total magnitude and that computed using our method for a simulated galaxy with a de Vaucouleurs light profile. A positive error means that the magnitude is overestimated (i.e., the luminosity is underestimated). The dashed line shows the sum of the other errors, which are mostly due to noise. For these simulations, galaxy scale lengths are taken from the fundamental plane for elliptical galaxies (Djorgovski & Davis 1987) and its extrapolation to faint magnitudes. (b) This figure shows why the luminosities of de Vaucouleurs galaxies are underestimated. The two lines are light profiles (without seeing) for galaxies with the same total luminosity inside $r = 2''$. Suppose that our detection brightness threshold is 0.7 in the units of the figure. We then measure identical $m_I$ and $r_1$ values for the two galaxies. Our method requires that we add the total light at $r > 2''$ for the exponential galaxy in order to get the total magnitude. This underestimates the luminosity of the de Vaucouleurs galaxy because it has a higher surface brightness than the exponential galaxy at all radii $r > 2''$.

**Figure 4.** A comparison between isophotal and aperture corrections and their uncertainties, for simulated exponential galaxies, assuming a night sky brightness of 21.0 mag arcsec$^{-2}$, poisson sky noise, and a seeing FWHM of 1.0 arcseconds. $\Delta m$ and $\sigma(\Delta m)$ are as defined in Figure 2. The short-dashed lines represent isophotal corrections (upper panel) and their uncertainties (lower panel). The long-dashed lines represent aperture corrections for a 3 arcsecond diameter aperture, and their uncertainties. The dotted-dashed lines represent aperture corrections for a 2 arcsecond diameter aperture, and their uncertainties. The thick lines represent the corrections for galaxies at $z = 0.02$ with apparent scale lengths computed from the scaling laws of giant late-type galaxies. The thin lines represent the corrections for galaxies at $z = 0.02$ with apparent scale lengths computed from the scaling laws of dSph galaxies.

**Figure 5.** The recovered $R$-band luminosity function for a simulated image using the method described in the text. Here $m$ is the recovered total apparent magnitude. The simulated image was for a cluster at $z \sim 0.2$ with no background or foreground galaxies. It consists of giant and dwarf galaxies with scale lengths given by the measured values for their local counterparts (Freeman 1970 for giants, Kormendy 1987b for dwarfs). All galaxies have exponential profiles; the pixel scale, sky brightness, and noise characteristics of the simulated image were typical for the $R$-band images listed in Table 2. The solid line is the true luminosity function. The points show the measured luminosity function; the error bars represent the quadrature sum of the uncertainties in the isophotal corrections [see Fig. 2(a)] and counting statistics.

**Figure 6.** Our measured $B$ and $R$ field counts, to be used as the background counts for our cluster fields. The $B$-band counts are averages for background fields 1 and 3 (see Table 2); the $R$-band counts are averages for background fields 1, 2, and 3. All counts have been corrected for Galactic extinction. The errors bars show the combined uncertainties due to the isophotal corrections and to counting statistics. In addition, a small contribution from uncertainties in the statistical completeness correction is included in the last point (at $m_L + 0.5$). Note that these errors represent errors in the **mean** counts; the standard deviation in the counts in an individual frame are approximately twice as big as these errors in both $B$ and $R$. The solid lines are the background counts from Driver et al. (1994b), and the dotted lines are the counts from Tyson (1988). The magnitudes here are total magnitudes; comparisons with published number counts that have been based on aperture magnitudes should be made with caution.

**Figure 7.** Predicted fluctuations $\left(\frac{\Delta N}{N}\right)_{\text{bkg}}$ in the background number counts for an image of area 46.5 square arcminutes, as appropriate for our $R$-band image of Field 1 of A194. The derivation of these curves is explained in Section 2(d) of the text.

**Figure 8.** Measured number counts (i.e., sum of cluster and background counts) for each of the fields listed in Table 2. The error bars are the quadrature sum of counting statistical errors and the uncertainties in aperture corrections. The dotted lines show the adopted mean background counts.

**Figure 9.** The luminosity function of Abell 194 (distance modulus = 34.30): (a) The $B$-band (open circles, dotted error bars) and $R$-band (filled circles, solid error bars) luminosity function for Field I. One galaxy in this field corresponds to log $N_{\rm gal}$ = 1.92; (b) The $R$-band luminosity function for Field II. One galaxy in this field corresponds to log $N_{\rm gal}$ = 1.84; (c) The isophotal magnitude ($m_I$) versus first-moment light radius ($r_1$, in units of arcseonds, as in Figure 2) correlation for objects in our $R$-band images of Fields I (filled circles), and II (open triangles). At a given $m_I$, lower surface-brightness galaxies have a higher $r_1$. The dashed line represents the line above which no background galaxies are observed in any of our background fields. The dotted-dashed line represent the line where we might expect typical dSph galaxies to lie, given the scaling laws described in the text and assuming a distance modulus of 34. The solid line represents the line above which exponential galaxies have too low surface-brightness to be detected with 100% statistical completeness. All objects in our catalog are presented (after the removal of spurious objects as described in Section 4). The narrow band of objects at low $r_1$ that extends to bright magnitudes are mostly stars; (d) The combined $R$-band luminosity function for Fields I and II. The total projected cluster area surveyed is $4.5 \times 10^4$ kpc$^2$.

**Figure 10.** The luminosity function of Abell 262 (distance modulus = 34.04): (a) The $B$-band (open circles, dotted error bars) and $R$-band (filled circles, solid error bars) luminosity function for Field I. One galaxy in this field corresponds to log $N_{\rm gal}$ = 1.94; (b) The $R$-band luminosity function for Field II. One galaxy in this field corresponds to log $N_{\rm gal}$ = 1.88; (c) The isophotal magnitude ($m_I$) versus first-moment light radius ($r_1$, in units of arcseconds) correlation for objects in our $R$-band images of Fields I (filled circles), and II (open triangles). The lines have the same meaning as in Figure 7(c); (d) The combined $R$-band luminosity function for Fields I and II. The total projected cluster area surveyed is $3.5 \times 10^4$ kpc$^2$.

**Figure 11.** The luminosity function of Pegasus (distance modulus = 33.75): (a) The $B$-band (open circles, dotted error bars) and $R$-band (filled circles, solid error bars) luminosity

function for Field I. One galaxy in this field corresponds to log $N_{\rm gal}$ = 1.97; (b) The $R$-band luminosity function for Field II. One galaxy in this field corresponds to log $N_{\rm gal}$ = 1.82; (c) The isophotal magnitude ($m_I$) versus first-moment light radius ($r_1$, in units arcseconds) correlation for objects in our $R$-band images of Fields I (filled circles), and II (open triangles). The lines have the same meaning as in Figure 7(c); (d) The combined $R$-band luminosity function for Fields I and II. The total projected cluster area surveyed is $2.6 \times 10^4$ kpc$^2$.

**Figure 12.** The luminosity function of the NGC 507 Group (distance modulus = 34.10): (a) The $B$-band (open circles, dotted error bars) and $R$-band (filled circles, solid error bars) luminosity function for Field I. One galaxy in this field corresponds to log $N_{\rm gal}$ = 1.95; (b) The $R$-band luminosity function for Field II. One galaxy in this field corresponds to log $N_{\rm gal}$ = 1.95; (c) The $R$-band luminosity function for Field III. One galaxy in this field corresponds to Log $N_{\rm gal}$ = 1.85; (d) The $R$-band luminosity function for Field IV. One galaxy in this field corresponds to Log $N_{\rm gal}$ = 1.84; (e) The isophotal magnitude ($m_I$) versus first-moment light radius ($r_1$, in units of arcseonds) correlation for objects in our $R$-band images of Fields I (filled circles), II (open squares), III (open hexagons), and IV (filled triangles). The lines have the same meaning as in Figure 7(c); (f) The combined $R$-band luminosity function for Fields I, II, III, and IV. The total projected cluster area surveyed is $6.9 \times 10^4$ kpc$^2$.

**Figure 13.** For A262 Field I, the upper panel shows the distribution of colors of galaxies with $20 < R < 23$ ($-14 < M_R < -11$ for galaxies in the cluster). Magnitudes were measured inside a $3''.0$ diameter aperture. The galaxies are bright enough so that the magnitudes are not significantly affected by random noise, and the aperture is large enough so that differential seeing between the $B$ and $R$ images is unimportant. Our simulations suggest that typical errors in the magnitudes for the galaxies represented here are, given their $r_1$ values, $\sim 0.04$ mag in $R$ and 0.06 mag in $B$. The typical error in each individual color is then 0.07 mag. The errors are larger for the faintest red galaxies ($R \sim 23$; $B - R > 2$), as these are only marginally detected in $B$. The histogram is 100 per cent complete for $B - R \leq 2.0$ assuming

that the apparent scale lengths of the galaxies are equal to those of local dwarfs moved to $z = 0.02$. The middle panel shows the analogous distribution for background field 1 in Table 2. The bottom panel shows the difference between the upper histograms; a correction for the difference in field area has been applied.

The bottom figure suggests that the excess cluster galaxies are red. Selection effects prevent us from finding the reddest galaxies; if these selection effects are severe, it would only make this statement stronger.

**Figure 14.** For NGC 507 Group Field I, the distribution of colors of galaxies with $20 < R < 23$ ($-14 < M_R < -11$ for galaxies in the cluster). The upper and lower panels are derived in the same way as the top and bottom panels of Figure 13. The uncertainties and completeness statistics are the same as for Figure 13.